\documentclass[prd,preprint,longbibliography,
  showpacs,showkeys,lengthcheck,
  nofootinbib,tightenlines,onecolumn,notitlepage,
  preprintnumbers,superscriptaddress]{revtex4-1}

\usepackage{graphicx}
\usepackage[usenames,dvipsnames]{xcolor}
\graphicspath{{figures/}}
\usepackage{tikz-feynman}

\usepackage{hyperref}
\hypersetup{colorlinks=true,citecolor=blue,linkcolor=blue,urlcolor=blue}

\usepackage{diagbox}
\usepackage{booktabs}

\usepackage{amsmath,amssymb,amsfonts}
\usepackage{mathrsfs}
\usepackage{slashed}
\usepackage{bm}
\usepackage{bbm}
\usepackage{mathtools}

\usepackage[draft]{changes} % Use for draft to show changes
\usepackage[utf8]{inputenc}
\usepackage{newtxtext}
\usepackage[mathcal]{euscript}

\renewcommand*{\d}{\mathop{}\!\mathrm{d}}
\newcommand*{\e}{\mathop{}\!\mathrm{e}}
\renewcommand*{\i}{\mathop{}\!\mathrm{i}}

\definecolor{darkgreen}{RGB}{0,120,0}

\newcommand*{\atan}{\mathop{}\!\mathrm{atan}}

\newcommand{\bd}{\bm{\varDelta}}
\newcommand{\hb}[1]{\hat{\bm{{#1}}}}
\newcommand{\lrn}{\overleftrightarrow{\nabla}}
\newcommand{\blrn}{\overleftrightarrow{\bm{\nabla}}}
\newcommand{\bla}{\mathfrak{A}}
\newcommand{\blt}{\mathfrak{t}}
\newcommand{\blc}{\bar{\mathfrak{c}}}
\newcommand{\blp}{\mathfrak{p}}
\newcommand{\blq}{\mathfrak{q}}

\newcommand{\psirel}{\psi}
\newcommand{\psibar}{\Psi}
\newcommand{\bohm}{\text{B}}

\begin{document}

\title{
  Quantum stresses in the hydrogen atom
}

\author{Adam Freese}
\email{afreese@jlab.org}
\address{Theory Center, Jefferson Lab, Newport News, Virginia 23606, USA}

\begin{abstract}
  Gravitational form factors are often interpreted
  as providing access to stresses inside hadrons, in particular
  through Fourier transforms of the form factors $D$ and $\bar{c}$.
  Some researchers, however, have expressed skepticism of this interpretation.
  I revisit the question,
  and argue that it is indeed appropriate to interpret
  these quantities as stress distributions.
  I consider the hydrogen atom's ground state as a familiar example,
  and use the pilot wave interpretation of quantum mechanics
  to give the distributions a clear meaning.
  A striking result is that $\bar{c}$---rather than $D$---quantifies the
  force law binding the system,
  which can be understood through Cauchy's first law of motion.
\end{abstract}

% Don't have this yet
\preprint{JLAB-THY-24-4243}

\maketitle

%%%%%%%%%%%%%%%%%%%%%%%%%%%%%%%%%%%%%%%%%%%%%%%%%%%%%%%%%%%%%%%%%%%%%%%%%%%%%%%%

\section{Introduction}
\label{sec:intro}

Mechanical properties of hadrons have received a lot of attention in recent
years~\cite{Polyakov:2018zvc,Burkert:2018bqq,Shanahan:2018nnv,Lorce:2018egm,Kumericki:2019ddg,Freese:2021czn,Burkert:2021ith,Panteleeva:2021iip,Pefkou:2021fni,Lorce:2021xku,Duran:2022xag,Burkert:2023wzr,Li:2024vgv}.
Form factors of the energy-momentum tensor (EMT)---which are commonly called
``gravitational form factors,''
owing to the EMT being the source of gravitation in general relativity---have
been interpreted as giving access to
spatial distributions of forces acting on quarks,
especially through the form factor often labeled $D$ and called the ``D-term.''
This culminated in a phenomenological extraction of pressures and shear forces
inside the proton by Burkert, Elouadrhiri and Girod
(BEG)~\cite{Burkert:2018bqq,Burkert:2021ith}.

However, skepticism about this interpretation of the D-term has been expressed,
especially by Ji and Liu~\cite{Ji:2021mfb},
who assert that the gravitational form factors instead provide access to
momentum current densities, and not to forces.
The contention can be understood by examining the continuity equation
for the energy-momentum tensor $T^{\mu\nu}(x)$,
which is locally conserved due to spacetime translation invariance of the
action~\cite{Kosyakov:2007qc,Leader:2013jra}:
\begin{align}
  \label{eqn:cont}
  \partial_\mu T^{\mu\nu}(x)
  =
  0
  \,.
\end{align}
If $n_\mu$ is a unit vector pointing in the forward time direction,
then
\begin{align}
  P^\nu(t)
  =
  \int \d^3 \bm{x} \,
  n_\mu T^{\mu\nu}(\bm{x},t)
\end{align}
is the total amount of four-momentum on a hypersurface of fixed $t$.
In integral form, Eq.~(\ref{eqn:cont}) can be restated:
\begin{align}
  \label{eqn:flux}
  \frac{\d}{\d t}
  \Big[
    P^\nu(V)
    \Big]
  =
  -
  \oint_{\partial V} \d \varSigma_i \,
  T^{i\nu}(\bm{x}, t)
  \,,
\end{align}
where $V$ is some spatial volume, $\partial V$ is its boundary,
and $P^\nu(V)$ the amount of four-momentum in the volume $V$.
This equation tells us that the amount of four-momentum in the region $V$
changes at the rate that its flux density, $T^{i\nu}(\bm{x},t)$,
enters or leaves the region through its boundary.
The components $T^{ij}$ in particular,
which are often called the stress tensor,
play the role of momentum flux densities.
This seems to validate the perspective of Ji and Liu.

However,
this does not entirely rule out an interpretation
in terms of local mechanical forces.
There are many classical systems for which the momentum flux density
clearly describes a pressure or the action of forces.
The most obvious example is a static fluid.
At a microscopic level,
the fluid consists of molecules that are in constant motion.
A body submerged in the fluid is pelted by the molecules at a high rate,
and from this experiences a pressure.
The momentum flux of moving particles is responsible for this pressure.

A deformable solid is another example.
Under the influence of an external force, the body will strain and deform.
The deformations will cause stress forces inside the body to act between
adjacent portions in an attempt to restore its original shape.
These forces will largely balance out---for a body in equilibrium,
the net force everywhere will be zero---but the individual forces
could actually be quite large.
In fact, if we divide a material body into two parts using
an open surface $\varSigma$,
the force exerted by one half on the other is given
by~\cite{chatterjee1999mathematical,chen2006meshless,irgens2008continuum,nayak2022continuum}:
\begin{align}
  \label{eqn:surface}
  F^j(\varSigma)
  =
  -
  \int_{\varSigma} \d \varSigma_i \,
  T^{ij}(\bm{x})
  \,,
\end{align}
which is exactly the momentum flux in Eq.~(\ref{eqn:flux})---but
now with an open surface.
It's quite easy to find scenarios in which the open surface integral of
Eq.~(\ref{eqn:surface}) can be quite large,
while the closed surface integral of Eq.~(\ref{eqn:flux}) can be small or even zero.
A static fluid is precisely such an example.
A bridge supporting traffic without collapsing is another.

It is not immediately clear that an analogy between
classical continuum materials and quarks in a proton is apt.
These are very different kinds of systems subject to different laws and dynamics.
Nonetheless, the comparison does raise the question:
does the stress tensor $T^{ij}$ in the proton describe the motion
of quarks, or local forces acting on quarks?
Perhaps it is even a mix of both.
Regardless of the answer, the gravitational form factor extracted by
BEG's analysis is telling us something very interesting about
what's happening inside a proton.
The question is: what exactly is it telling us?

The question is difficult to answer for many reasons.
The appropriate definition of spatial densities in
relativistic quantum field theory has itself been a subject of controversy,
with multiple competing frameworks~\cite{Sachs:1962zzc,Fleming:1974af,Burkardt:2000za,Diehl:2000xz,Burkardt:2002hr,Miller:2007uy,Miller:2018ybm,Lorce:2020onh,Jaffe:2020ebz,Freese:2021czn,Panteleeva:2021iip,Freese:2021mzg,Epelbaum:2022fjc,Li:2022ldb,Panteleeva:2022khw,Freese:2022fat,Panteleeva:2022uii,Chen:2022smg,Freese:2023jcp,Panteleeva:2023evj,Freese:2023abr,Li:2024vgv}.
The operators in question also contain ultraviolet divergences
that must be renormalized,
and subtleties about the renormalization rules for composite operators
have also led to controversy~\cite{Lorce:2021xku}.

There is an even more fundamental difficulty
which hasn't received as much attention.
Quantum chromodynamics is a quantum theory, and quantum mechanics
on its own doesn't have a clear ontology---i.e.,
it doesn't tell a clear story about what actually happens
in microscopic systems.
Vanilla quantum mechanics is in essence a set of tools
for calculating expectation values and event rates,
and does not paint a vivid picture of what is actually going on
in the microscopic world of quarks and gluons.
It is thus unclear when (or even if)
we can say things like
``a quark located at $\bm{x}$ feels a pressure $p$ and has a velocity $\bm{v}$,''
even statistically.
The Copenhagen interpretation would even tell
us that these statements are outright meaningless,
and that that quantum mechanics is \emph{only} a set of tools
for making predictions for concrete experimental
scenarios~\cite{Bohr:1937cc,Bohr:2017atm}.

With all of these issues piling on,
it is prudent to consider simplified cases and to try to address the
issues one at a time.
My goal in this paper is to address the question of how
the expectation value of the stress tensor could be
interpreted in quantum mechanics,
if it has been supplemented with a clear ontology.
To avoid the dogpile of issues related to relativity and field theory,
and to keep the problem as simple as possible,
I will strictly limit my attention to the ground state of the hydrogen atom
(which was previously been considered by Ji and Liu~\cite{Ji:2022exr}).

As an ontology for quantum mechanics,
I consider the pilot wave
interpretation~\cite{db:pilot,Bohm:1951xw,Bohm:1951xx,debroglie1960non,Bell:2004gpx,Bohm:2006und},
which is often also called Bohmian mechanics.
In the pilot wave interpretation,
every particle in a system has a definite position at all times.
Additionally, the wave function is considered to be a
physically real field---albeit, living in configuration space
rather than the three-dimensional space of our experiences---which
guides the motion of the particles.
The wave function also provides, as usual,
the probability of the system having a particular configuration
at any time via Born's rule.
The utility of this formulation is that with exact trajectories,
it is crystal clear what we mean
when we say that a particle has a force acting on it or that it is in motion.
We merely need to take an ensemble average of all possible configurations
via Born's rule, due to our ignorance of the exact configuration.

The paper is divided into three main sections.
First, in Sec.~\ref{sec:gff},
I obtain the gravitational form factors of the hydrogen atom's ground state.
Next, in Sec.~\ref{sec:space},
I obtain impact parameter space expressions through 3D Fourier transforms.
These two sections are vanilla quantum mechanics,
and the results therein do not rely on any particular interpretation.
In Sec.~\ref{sec:bohm}, I use the pilot wave formulation
to interpret the impact parameter expressions of Sec.~\ref{sec:space}.
I argue that the stress tensor should be taken to provide literal
stresses inside the atom,
finding that the particle contributions in particular
correspond to forces exerted by the pilot wave on the electron and proton.
Additionally, I find through Cauchy's first law of motion that
individual contributions to the form factor $\bar{c}(\bd^2)$---rather
than to $D(\bd^2)$---encode the force law binding the system.
I finally summarize and provide an outlook in Sec.~\ref{sec:end}.

%%%%%%%%%%%%%%%%%%%%%%%%%%%%%%%%%%%%%%%%%%%%%%%%%%%%%%%%%%%%%%%%%%%%%%%%%%%%%%%%

\section{Gravitational form factors}
\label{sec:gff}

In the non-relativistic regime, the most general form that matrix elements
of the stress tensor can take for a system without spin---owing to
Galilean covariance, and time reversal and spatial inversion
symmetries---is\footnote{
  Using natural units: $\hslash=c=1$.
}:
\begin{align}
  \label{eqn:gff}
  \langle \bm{p}' | \hat{T}^{ij}(0) | \bm{p} \rangle
  =
  \frac{P^i P^j}{M}
  A(\bd^2)
  +
  \frac{\varDelta^i \varDelta^j - \bd^2 \delta^{ij}}{4M}
  D(\bd^2)
  -
  M \delta^{ij} \bar{c}(\bd^2)
  \,,
\end{align}
where $M$ is the mass of the composite system,
$\bm{p}$ and $\bm{p}'$ are (initial and final) barycentric momenta,
$\bm{P} = \frac{1}{2}\big( \bm{p} + \bm{p}' \big)$,
and
$\bd = \bm{p}' - \bm{p}$,
The operator $\hat{T}^{ij}(\bm{x})$ is the stress tensor operator
in the Schr\"{o}dinger picture,
and $A(\bd^2)$, $D(\bd^2)$ and
$\bar{c}(\bd^2)$ are the gravitational form factors.
I have written Eq.~(\ref{eqn:gff}) so that the gravitational form factors
match those of the standard decomposition~\cite{Polyakov:2018zvc}
in the non-relativistic limit.

An important sum rule that follows from the continuity equation (\ref{eqn:cont})
is $\bar{c}(\bd^2)=0$.
However,
contributions to $\bar{c}(\bd^2)$ from individual constituents
(in this case, the electron, proton and Coulomb field)
may be non-zero,
and these separate contributions are crucial to
the physical meaning of the stress tensor.
After all, even in familiar classical systems such as a bridge carrying traffic,
stresses are individual forces that balance out.
It is thus prudent to keep track of $\bar{c}(\bd^2)$.

For a composite system, $|\bm{p}\rangle$ in Eq.~(\ref{eqn:gff}) should be
understood to implicitly contain some state for the internal structure of
the system.
For instance,
the non-relativistic hydrogen atom is characterized by its quantum numbers
$n,l,m_l$, in addition to a barycentric position $\bm{r}$ or momentum $\bm{p}$.
Thus,
\begin{align}
  |\bm{p}\rangle \equiv |\bm{p};n,l,m_l\rangle
\end{align}
for some set of quantum numbers.
The matrix element in Eq.~(\ref{eqn:gff}) will depend on which
internal state $|n,l,m_l\rangle$ is used;
different states of the hydrogen atom will have different form factors.
Of course, only $l=0$ states will satisfy Eq.~(\ref{eqn:gff}),
since the breakdown therein assumes a system without spin.
While it would be interesting to look at the stress tensor for $l>0$ states,
and thus study the implications of orbital angular momentum for the stress tensor,
I limit the present study to $l=0$ for simplicity.

%%%%%%%%%%%%%%%%%%%%%%%%%%%%%%%%%%%%%%%%

\subsection{The quantum stress tensor}

The stress tensor operator is needed to evaluate
the matrix element in Eq.~(\ref{eqn:gff}).
The operator can be broken down into contributions from the
particles (electron and proton) and from the Coulomb field:
\begin{align}
  \hat{T}^{ij}(\bm{x})
  &=
  \sum_{a=e,p,\gamma}
  \hat{T}_a^{ij}(\bm{x})
  \,.
\end{align}
The particle stress operator should be a quantization of
the classical stress tensor for a free particle:
\begin{align}
  \label{eqn:stress:cl}
  T^{ij}_{\mathrm{cl.}}(\bm{x},t)
  =
  \frac{p^i(t) p^j(t)}{m}
  \delta^{(3)}(\bm{x} - \bm{q}(t))
  \,,
\end{align}
where $\bm{q}(t)$ is the particle's trajectory and $\bm{p}(t)$ its momentum.
Unfortunately, first quantization is an ambiguous procedure;
there isn't a unique scheme to assign a corresponding quantum operator.
The issue is that different operators, such as:
\begin{align*}
  \hat{T}^{ij}(\bm{x})
  &\stackrel{?}{=}
  \frac{1}{2m}
  \Big(
  \hat{p}^i
  \delta^{(3)}(\bm{x} - \hb{q})
  \hat{p}^j
  +
  \hat{p}^j
  \delta^{(3)}(\bm{x} - \hb{q})
  \hat{p}^i
  \Big)
  \\
  \hat{T}^{ij}(\bm{x})
  &\stackrel{?}{=}
  \frac{1}{2m}
  \Big(
  \hat{p}^i
  \hat{p}^j
  \delta^{(3)}(\bm{x} - \hb{q})
  +
  \delta^{(3)}(\bm{x} - \hb{q})
  \hat{p}^i
  \hat{p}^j
  \Big)
  \,,
\end{align*}
among many others,
all correspond to the same classical quantity.
To be sure, this ambiguity is not present when second quantizing field theories,
but we must navigate this obstacle for the goal at hand.

Although I am otherwise avoiding field theory in this paper,
known results from quantum field theory can help to pick the appropriate operator.
In free field theories, the form factors
$A(\bd^2)$ and $D(\bd^2)$ are constants
and $\bar{c}(\bd^2)=0$.
In particular, $A(\bd^2) = 1$,
while the value of $D(\bd^2)$ depends on the particle species in question.
For free fermions, $D(\bd^2)=0$~\cite{Hudson:2017oul},
while for scalar particles obeying the Klein-Gordon equation, $D(\bd^2)=-1$~\cite{Hudson:2017xug}.
Since matter particles are typically fermions,
I will choose a stress tensor operator that gives $A(\bd^2)=1$
and $D(\bd^2)=0$ for single-particle states.
Even if $D(\bd^2)$ were not zero for a free particle,
its value is a red herring that will lead us off the trail
of understanding what these form factors tell us about bound states.

The appropriate operator for the particle stress tensor is:
\begin{align}
  \label{eqn:stress:op}
  \hat{T}^{ij}(\bm{x})
  =
  \frac{1}{4m}
  \Big\{
    \hat{p}^i ,
    \big\{
      \hat{p}^j ,
      \delta^{(3)}(\bm{x}-\hb{q})
      \big\}
    \Big\}
  \,,
\end{align}
where $\{a,b\} = ab + ba$ is the anticommutator.
According to McCoy's formula~\cite{McCoy:1929com},
this is the Weyl quantized operator associated with
the classical stress tensor (\ref{eqn:stress:cl}).
Sandwiching the operator (\ref{eqn:stress:op}) between momentum kets gives:
\begin{align}
  \langle \bm{p}' | \hat{T}^{ij}(0) | \bm{p} \rangle
  =
  \frac{P^i P^j}{m}
  \,,
\end{align}
where $\bm{P} = \frac{1}{2} \big(\bm{p}+\bm{p}'\big)$.
Comparison with the form factor breakdown (\ref{eqn:gff})
tells us that $A(\bd^2)=1$ and $D(\bd^2)=\bar{c}(\bd^2)=0$, as required.

In the presence of general electromagnetic fields,
the operator (\ref{eqn:stress:op}) should be modified via minimal substitution
in the usual way, with $\hat{p}^i \mapsto \hat{p}^i - e \hat{A}^i(\bm{x},t)$.
For the hydrogen atom---at least in the absence of
fine structure corrections---the magnetic field is zero,
and thus one can choose a gauge where $\bm{A} = 0$.
I will exploit that freedom throughout this work.
In the $\bm{A}=0$ gauge, when taking matrix elements,
the momentum operator becomes $-\i\bm{\nabla}$
acting on the relevant wave packets.
In a more general gauge, $\bm{\nabla}$ is simply replaced---via
minimal substitution---by the gauge-covariant derivative.
Accordingly, the matrix elements of the operator (\ref{eqn:stress:op})
are gauge-invariant after minimal substitution has been done.
Therefore, although I use the $\bm{A}=0$ gauge throughout this work
for simplicity, the results are all gauge-invariant.

In addition to the electron and proton,
the electric field can also carry stresses.
The formula for the electromagnetic energy-momentum tensor is:
\begin{align}
  \hat{T}_\gamma^{\mu\nu}(x)
  =
  \hat{F}^{\mu\rho}(x) \hat{F}_{\rho}^{\phantom{\rho}\nu}(x)
  +
  \frac{1}{4} g^{\mu\nu}
  \hat{F}_{\alpha\beta}(x)
  \hat{F}^{\alpha\beta}(x)
  \,.
\end{align}
The stress tensor (spacelike components) for a static electric field reduces to:
\begin{align}
  \label{eqn:Tg:op}
  \hat{T}_\gamma^{ij}(\bm{x})
  =
  -\hat{E}^i(\bm{x})
  \hat{E}^j(\bm{x})
  +
  \frac{1}{2} \delta^{ij}
  \big(\hb{E}(\bm{x})\big)^2
  \,.
\end{align}
The electric field in the hydrogen atom
is a sum of two static Coulomb fields from point sources
at positions $\hb{q}_e$ and $\hb{q}_p$.
Crucially, since this is a quantum mechanical system,
these positions are operators.

%%%%%%%%%%%%%%%%%%%%%%%%%%%%%%%%%%%%%%%%

\subsection{Particle form factors}

With the appropriate operators in hand,
gravitational form factors can be obtained.
I start with the particle contributions.
After the mass, momentum and position have been given the subscript $a\in\{e,p\}$,
sandwiching the operator (\ref{eqn:stress:op}) between
momentum kets gives:
\begin{align}
  \langle \bm{p}_e', \bm{p}_p' |
  \hat{T}^{ij}_{a}(\bm{x})
  | \bm{p}_e, \bm{p}_p \rangle
  =
  \frac{P^i_{a} P^j_{a}}{m_{a}}
  \e^{-\i \bd\cdot\bm{x}}
  (2\pi)^3 \delta^{(3)}(\bd - \bd_{a})
  \,,
\end{align}
where
$\bm{P}_a = \frac{1}{2}\big( \bm{p}_a + \bm{p}'_a \big)$
and
$\bd_a = \bm{p}'_a - \bm{p}_a$
are per-particle momenta,
and where
$\bd = \bd_e + \bd_p$ and $\bm{P} = \bm{P}_e + \bm{P}_p$
are total momenta.
I transform to barycentric coordinates to proceed.
The usual formula
\begin{align}
  \notag
  \bm{p}
  &=
  \bm{p}_e + \bm{p}_p
  \\
  \bm{k}
  &=
  \frac{m_p \bm{p}_e - m_e \bm{p}_p}{m_e+m_p}
  \,,
\end{align}
is applied to initial (unprimed), final (primed)
and total (capitalized) momenta alike.
As usual,
\begin{align}
  \notag
  M &= m_e + m_p
  \\
  \mu &= \frac{m_e m_p}{m_e + m_p}
\end{align}
are the total and reduced masses.
In terms of the transformed variables,
the individual stress tensor matrix elements can be written:
\begin{align}
  \langle \bm{p}'; \bm{k}' |
  \hat{T}_a^{ij}(\bm{x})
  | \bm{p}; \bm{k} \rangle
  &=
  \left(
  \frac{m_a P^i P^j}{M^2}
  \pm
  \frac{P^i K^j + K^i P^j}{M}
  +
  \frac{K^i K^j}{m_a}
  \right)
  \e^{-\i \bd\cdot\bm{x}}
  (2\pi)^3 \delta^{(3)}\left(\pm\frac{\mu}{m_a}\bd - \bm{k}' + \bm{k}\right)
  \,,
\end{align}
where here (and elsewhere) the top of $\pm$ applies for $a=e$
and the bottom for $a=p$.
Now, to perform a form factor decomposition (\ref{eqn:gff}),
the same operator should be sandwiched
between kets with a definite stationary
state in place of the relative momentum,
e.g., $\langle \bm{p}'; \psirel(t) | \ldots | \bm{p}; \psirel(t)\rangle$
for some internal state $|\psirel(t)\rangle$.
If $\psirel(\bm{r},t)$ is the wave function in configuration space
(with $\bm{r} = \bm{q}_e - \bm{q}_p$ being the relative separation), then:
\begin{align}
  \label{eqn:stress:chi}
  \langle \bm{p}'; \psirel(t) |
  \hat{T}_a^{ij}(0)
  | \bm{p}; \psirel(t) \rangle
  &=
  \int \d^3 \bm{r} \,
  \e^{\pm\i\left(\frac{\mu\bd}{m_a}\right)\cdot\bm{r}}
  \psirel^*(\bm{r},t)
  \left(
  \frac{m_a P^i P^j}{M^2}
  \mp
  \i
  \frac{P^i \overleftrightarrow{\nabla}^j + \overleftrightarrow{\nabla}^i P^j}{2M}
  -
  \frac{\overleftrightarrow{\nabla}^i \overleftrightarrow{\nabla}^j}{4m_a}
  \right)
  \psirel(\bm{r},t)
  \,,
\end{align}
where the two-sided derivative is defined as
$f \lrn g = f(\nabla g) - (\nabla f)g$.
This expression so far applies to general two-body systems,
and can be used as a departing point for future work
other states and systems.
However, I will presently specialize to the hydrogen atom's ground state.

The formula for the hydrogen atom's ground state wave function is well-known:
\begin{align}
  \label{eqn:wf}
  \psirel(\bm{r})
  =
  \sqrt{\frac{(\alpha \mu)^3}{\pi}}
  \e^{-\alpha \mu |\bm{r}|}
  \,,
\end{align}
where $\alpha = \frac{e^2}{4\pi}$ is the fine structure constant.
Since this wave function is real,
the second structure appearing in Eq.~(\ref{eqn:stress:chi}),
with a factor $P^i \lrn^j + \lrn^i P^j$,
vanishes.
Of the remaining structures,
the tensor structure associated with $A(\bd^2)$
it is that with a factor $P^i P^j/M$.
The remaining $\lrn^i\lrn^j$ term is associated with the remaining form factors
$D(\bd^2)$ and $\bar{c}(\bd^2)$.
To separate them, it is helpful to notice that the tensor structure
associated with $D(\bd^2)$ in Eq.~(\ref{eqn:gff}) is orthogonal to $\bd$.
Thus, $\bar{c}(\bd^2)$ can be isolated by contracting this expression with
$-\frac{1}{M}\frac{\varDelta^i \varDelta^j}{\bd^2}$.
The remaining $D(\bd^2)$ form factor is then obtained by subtracting the
$\bar{c}(\bd^2)$ part off.
This produces the following formulas for the form factors:
\begin{align}
  \label{eqn:A}
  A_a(\bd^2)
  &=
  \frac{m_a}{M}
  \int \d^3 \bm{r} \,
  \e^{\pm\i\left(\frac{\mu\bd}{m_a}\right)\cdot\bm{r}}
  \psirel^*(\bm{r},t)
  \psirel(\bm{r},t)
  \\
  \label{eqn:D}
  D_a(\bd^2)
  &=
  \frac{M}{2 m_a \bd^2}
  \int \d^3 \bm{r} \,
  \e^{\pm\i\left(\frac{\mu\bd}{m_a}\right)\cdot\bm{r}}
  \psirel^*(\bm{r})
  \blrn^2
  \psirel(\bm{r})
  -
  \frac{6 M^2}{\bd^2}
  \bar{c}_a(\bd^2)
  \\
  \label{eqn:cbar}
  \bar{c}_a(\bd^2)
  &=
  \frac{1}{4 m_a M \bd^2}
  \int \d^3 \bm{r} \,
  \e^{\pm\i\left(\frac{\mu\bd}{m_a}\right)\cdot\bm{r}}
  \psirel^*(\bm{r},t)
  \big(\bd\cdot\blrn\big)^2
  \psirel(\bm{r},t)
  \,.
\end{align}
Plugging the wave function (\ref{eqn:wf}) into these expressions
gives the following results:
\begin{align}
  \label{eqn:A:ep}
  A_a(\bd^2)
  &=
  \frac{m_a}{M}
  \frac{ 1 }{ (1 + \tau_a)^2 }
  \\
  \label{eqn:D:ep}
  D_a(\bd^2)
  &=
  -
  \frac{M \mu^2}{2m_a^3\tau_a}
  \left\{
    \frac{
      1
    }{
      1 + \tau_a
    }
    -
    \frac{3}{\tau_a}
    \left[
      1
      -
      \frac{1}{\sqrt{\tau_a}}
      \atan(\sqrt{\tau_a})
      \right]
    \right\}
  \\
  \label{eqn:cbar:ep}
  \bar{c}_a(\bd^2)
  &=
  -
  \frac{\alpha^2\mu^2}{Mm_a\tau_a}
  \left[
    1
    -
    \frac{1}{\sqrt{\tau_a}}
    \atan(\sqrt{\tau_a})
    \right]
  \,,
\end{align}
where
\begin{align}
  \label{eqn:tau}
  \tau_a
  &=
  \frac{\bd^2}{(2\alpha m_a)^2}
  \,.
\end{align}
It is worth remarking that
$D_a(0)$ and $\bar{c}_a(0)$ are both finite:
\begin{align}
  D_a(0)
  &=
  \frac{\mu^2 M}{5m_a^3}
  \\
  \bar{c}_a(0)
  &=
  -
  \frac{\alpha^2 \mu^2}{3 M m_a}
  \,.
\end{align}

%%%%%%%%%%%%%%%%%%%%%%%%%%%%%%%%%%%%%%%%

\subsection{Coulomb field form factors}

The contributions of the Coulomb field
are found next by sandwiching
the operator (\ref{eqn:Tg:op})
between momentum kets.
The expression in Eq.~(\ref{eqn:Tg:op}) is gauge-invariant,
but for convenience I use the $\bm{A}=0$ gauge,
so that the electric field can be written in terms of a scalar potential via
$\hb{E}(\bm{x}) = -\bm{\nabla} \hat{\Phi}(\bm{x})$,
the latter of which is a sum of point source potentials:
\begin{align}
  \hat{\Phi}(\bm{x})
  =
  \sum_{a=e,p}
  \frac{e_a}{4\pi|\bm{x} - \hb{q}_a|}
  \,.
\end{align}
To perform momentum-space calculations,
it is convenient to write the potential in terms of a Fourier transform.
I will use a trick to relate this potential to a Yukawa potential in the
zero mass limit, since it has an easy-to-use momentum space expression:
\begin{align}
  \hat{\Phi}(\bm{x})
  =
  \lim_{m_\gamma\rightarrow0}
  \sum_{a=e,p}
  \frac{e_a}{4\pi|\bm{x} - \hb{q}_a|}
  \e^{-m_\gamma |\bm{x} - \hb{q}_a|}
  =
  \lim_{m_\gamma\rightarrow0}
  \sum_{a=e,p}
  e_a
  \int \frac{\d^3\bm{k}}{(2\pi)^3}
  \frac{\e^{\i \bm{k}\cdot(\bm{x}-\hb{q}_a)}}{\bm{k}^2 + m_\gamma^2}
  \,.
\end{align}
Using this Fourier transform in Eq.~(\ref{eqn:Tg:op}),
and sandwiching between momentum kets, gives:
\begin{multline}
  \langle \bm{p}' |
  \hat{T}_\gamma^{ij}(0)
  | \bm{p} \rangle
  =
  -\frac{e^2}{2}
  \lim_{m_\gamma \rightarrow 0}
  \int \d^3 \bm{r} \,
  \big| \psirel(\bm{r}) \big|^2
  \int \frac{\d^3\bm{k}}{(2\pi)^3}
  \frac{
    2 k^i k^j - \frac{1}{2} \varDelta^i \varDelta^j
    -
    \delta^{ij} \bm{k}^2
    +
    \frac{1}{4} \delta^{ij} \bd^2
  }{
    \left[\left(\bm{k} + \frac{1}{2}\bd\right)^2 + m_\gamma^2\right]^2
    \left[\left(\bm{k} - \frac{1}{2}\bd\right)^2 + m_\gamma^2\right]^2
  }
  \Bigg\{
    \exp\left(\i \frac{\mu}{m_e} \bd\cdot\bm{r}\right)
    \\
    +
    \exp\left(-\i \frac{\mu}{m_p} \bd\cdot\bm{r}\right)
    -
    \exp\left(\i\left[\bm{k} + \frac{m_p-m_e}{2M} \bd\right]\cdot\bm{r}\right)
    -
    \exp\left(-\i\left[\bm{k} + \frac{m_p-m_e}{2M} \bd\right]\cdot\bm{r}\right)
    \Bigg\}
  \,.
\end{multline}
It is safe to take the $m_\gamma \rightarrow 0$ limit.
The integral over $\bm{r}$
is the same as that going from Eq.~(\ref{eqn:A}) to (\ref{eqn:A:ep}),
and gives:
\begin{multline}
  \langle \bm{p}' |
  \hat{T}_\gamma^{ij}(0)
  | \bm{p} \rangle
  =
  -\frac{e^2}{2}
  \int \frac{\d^3\bm{k}}{(2\pi)^3}
  \frac{
    2 k^i k^j - \frac{1}{2} \varDelta^i \varDelta^j
    -
    \delta^{ij} \bm{k}^2
    +
    \frac{1}{4} \delta^{ij} \bd^2
  }{
    \left(\bm{k} + \frac{1}{2}\bd\right)^2
    \left(\bm{k} - \frac{1}{2}\bd\right)^2
  }
  \Bigg\{
    \frac{1}{(1+\tau_e)^2}
    \\
    +
    \frac{1}{(1+\tau_p)^2}
    -
    \frac{2(2\alpha\mu)^4}{
      \left(
      (2\alpha\mu)^2 + \big[ \bm{k} + \frac{m_p-m_e}{2M} \bd \big]^2
      \right)^2
    }
    \Bigg\}
  \,.
\end{multline}
The remaining $\bm{k}$ integrals can be evaluated
using standard Feynman parametrization techniques.
The resulting form factors are:
\begin{align}
  \label{eqn:A:g}
  A_\gamma(\bd^2)
  &=
  0
  \\
  \label{eqn:D:g}
  D_\gamma(\bd^2)
  &=
  \frac{\alpha\pi M}{4|\bd|}
  \left[
    \frac{1}{(1+\tau_e)^2}
    +
    \frac{1}{(1+\tau_p)^2}
    \right]
  +
  \frac{4\alpha^3 \mu^2 M}{|\bd|^3}
  \frac{1}{\nu^2}
  \left[
    1 - \frac{\atan(\nu)}{\nu}
    \right]
  -
  \frac{6M^2}{\bd^2} \bar{c}_\gamma(\bd^2)
  \\
  \label{eqn:cbar:g}
  \bar{c}_\gamma(\bd^2)
  &=
  \frac{\alpha^2 \mu^2}{M}
  \sum_{a=e,p}
  \frac{1}{m_a \tau_a}
  \left[
    1
    -
    \frac{1}{\sqrt{\tau_a}}
    \atan\big(\sqrt{\tau_a}\big)
    \right]
  \,,
\end{align}
where
\begin{align}
  \nu
  =
  \frac{
    1 - \sqrt{\tau_e \tau_p}
  }{
    \sqrt{\tau_e} + \sqrt{\tau_p}
  }
\end{align}
and where $\tau_a$ is defined in Eq.~(\ref{eqn:tau}).
As with the particle case, both $D_\gamma(0)$
and $\bar{c}_\gamma(0)$ are finite:
\begin{align}
  D_\gamma(0)
  &=
  \frac{4}{5}
  \left(
  \frac{M}{\mu}
  -
  \frac{1}{2}
  \right)
  \\
  \bar{c}_\gamma(0)
  &=
  \frac{\alpha^2 \mu}{3 M}
  \,.
\end{align}

%%%%%%%%%%%%%%%%%%%%%%%%%%%%%%%%%%%%%%%%

\subsection{Discussion and numerical results}

Let us briefly examine the results so far before moving on.
There are two sum rules that must be satisfied.
The first is usually called the momentum sum rule:
\begin{align}
  A_e(0)
  +
  A_p(0)
  +
  A_\gamma(0)
  =
  1
  \,,
\end{align}
which---examining Eqs.~(\ref{eqn:A:ep}) and (\ref{eqn:A:g})---is satisfied.
Additionally, local momentum conservation imposes the following sum rule:
\begin{align}
  \bar{c}_e(\bd^2)
  +
  \bar{c}_p(\bd^2)
  +
  \bar{c}_\gamma(\bd^2)
  =
  0
  \,,
\end{align}
which---examining Eqs.~(\ref{eqn:cbar:ep}) and (\ref{eqn:cbar:g})---is
also satisfied.
This is somewhat remarkable: although the calculation is non-relativistic,
momentum is apparently locally rather than instantaneously transferred
between the particles.

Next, let us compare to existing works.
Ji and Liu~\cite{Ji:2022exr} previously gave results for the form factor
$C(\bd^2) = \frac{m_e}{4M} D(\bd^2)$.
The bulk of their paper involves
higher-order radiative corrections to this form factor,
but leading order results are given in Sec.~II.C.
Ji and Liu in effect use an infinite proton mass
(they assume the proton to be stationary at the origin),
so my results should reproduce theirs in that limit.
Taking $m_p \rightarrow\infty$ in
Eqs.~(\ref{eqn:D:ep}) and (\ref{eqn:D:g}) gives:
\begin{align}
  \notag
  C_e(\bd^2)
  & \xrightarrow[m_p\rightarrow\infty]{}
  \frac{1}{8}
  \left\{
    -\frac{1}{\tau_e}
    \frac{1}{1+\tau_e}
    +
    \frac{3}{\tau_e^2}
    \left[
      1 - \frac{1}{\sqrt{\tau_e}} \atan(\sqrt{\tau_e})
      \right]
    \right\}
  \\
  \notag
  C_p(\bd^2)
  & \xrightarrow[m_p\rightarrow\infty]{}
  0
  \\
  \notag
  C_\gamma(\bd^2)
  & \xrightarrow[m_p\rightarrow\infty]{}
  \frac{1}{8}
  \left\{
    \frac{1}{\tau_e}
    +
    \frac{1}{\sqrt{\tau_e}}
    \atan(\sqrt{\tau_e})
    -
    \frac{
      \pi\sqrt{\tau_e}
      (\tau_e+2)
    }{
      4
      (1+\tau_e)^2
    }
    -
    \frac{3}{\tau_e^2}
    \left[
      1
      -
      \frac{1}{\sqrt{\tau_e}}
      \atan\big(\sqrt{\tau_e}\big)
      \right]
    \right\}
  \\
  C(\bd^2)
  =
  \sum_{a=e,p,\gamma} C_a(\bd^2)
  & \xrightarrow[m_p\rightarrow\infty]{}
  \frac{1}{8}
  \left\{
    \frac{1}{1+\tau_e}
    +
    \frac{1}{\sqrt{\tau_e}}
    \atan(\sqrt{\tau_e})
    -
    \frac{
      \pi\sqrt{\tau_e}
      (\tau_e+2)
    }{
      4
      (1+\tau_e)^2
    }
    \right\}
  \,,
\end{align}
which agrees with Ji and Liu's results,
if their Eqs.~(56), (57) and (59) are summed.

Unlike $D(\bd^2)$, Ji and Liu did not give results for $A(\bd^2)$ nor $\bar{c}(\bd^2)$.
The calculation of these form factors for the hydrogen ground state
is new to this work.

A significant finding of Ji and Liu is that $C(0) = \frac{1}{4} > 0$.
I reproduce their finding, with:
\begin{align}
  D(0)
  =
  \sum_{a=e,p,\gamma}
  D_a(0)
  =
  \frac{M}{\mu}
  -
  1
  \,.
\end{align}
The smallest value $D(0)$ can take is $3$ in the mass-balanced case.
For highly mass-imbalanced cases, it grows significantly;
$D(0) \approx 1837$ if the physical proton and electron masses are used.

It has often been speculated
(see Refs.~\cite{Polyakov:2018zvc,Lorce:2018egm,Freese:2021czn} for instance)
that $D(0) < 0$ is a necessary mechanical stability criterion for composite systems.
Since the hydrogen atom's ground state is stable,
I agree with Ji and Liu that this counterexample refutes the negativity conjecture,
and that the sign of $D(0)$ must have nothing to do with mechanical stability.

\begin{figure}
  \includegraphics[width=0.33\textwidth]{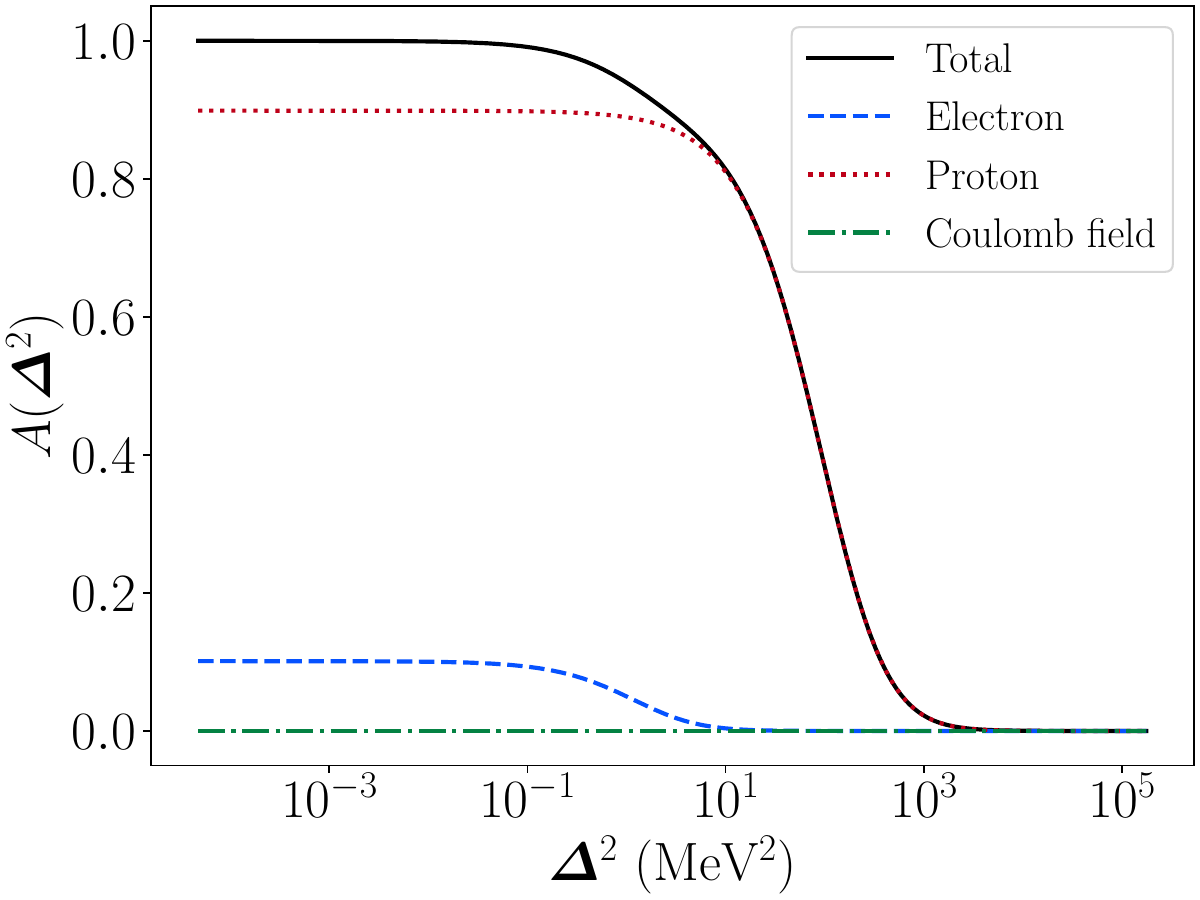}
  \includegraphics[width=0.33\textwidth]{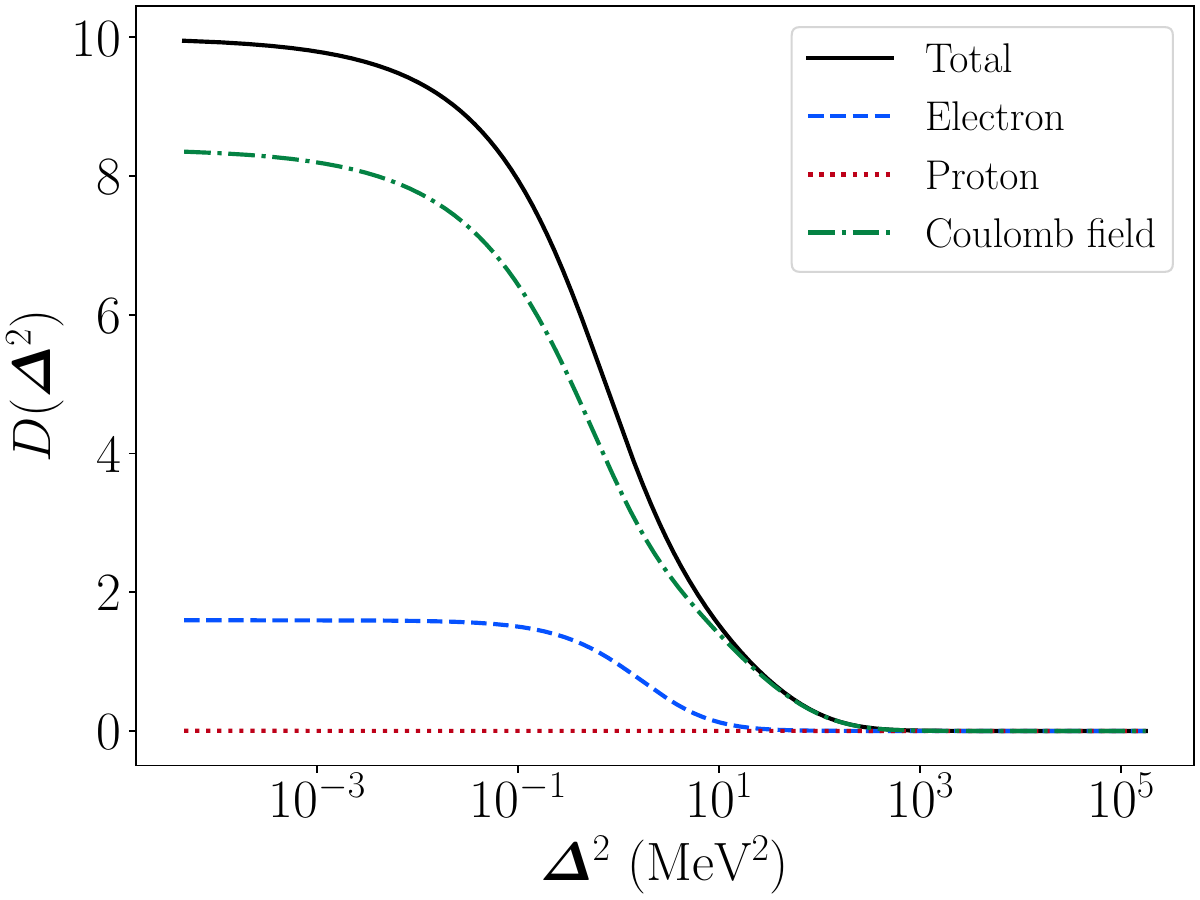}
  \includegraphics[width=0.33\textwidth]{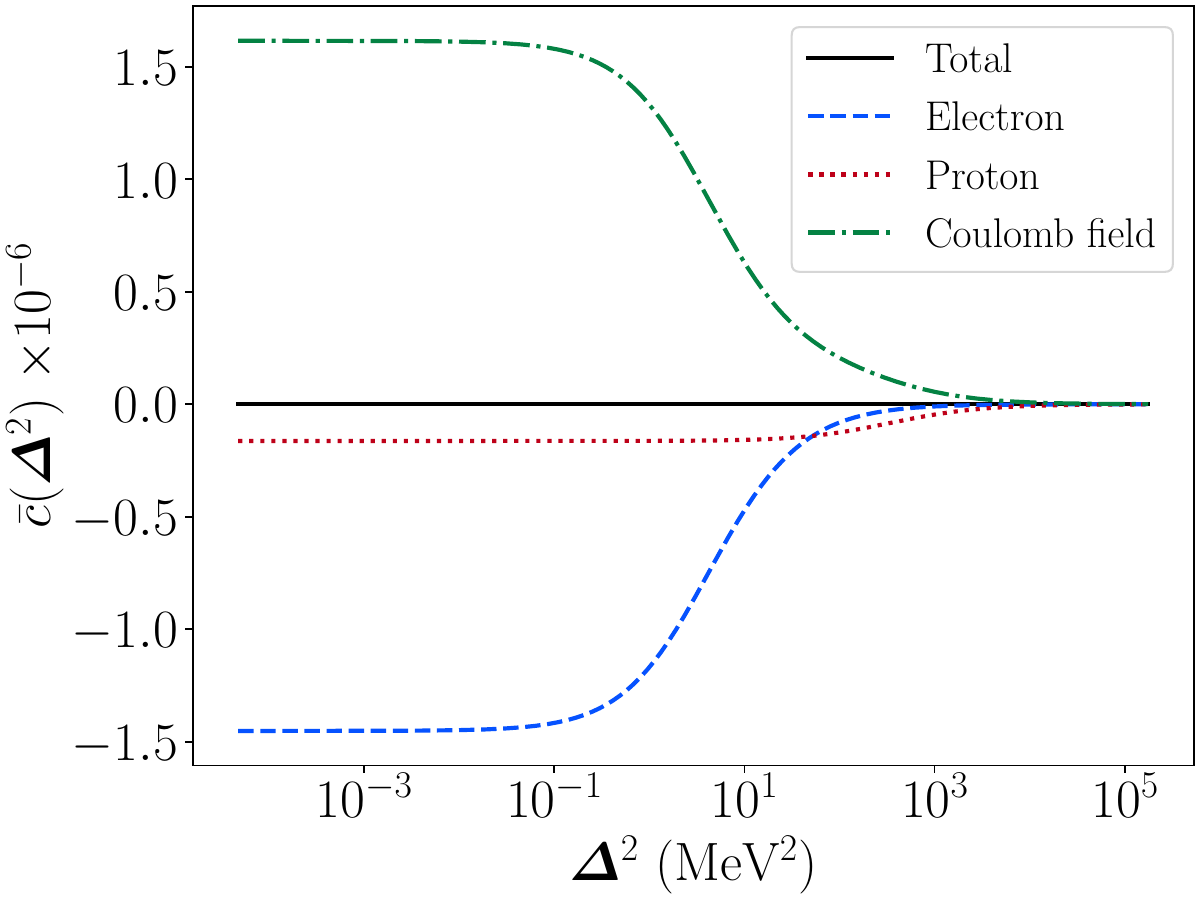}
  \caption{
    Plots of the gravitational form factors of the hydrogen atom in its
    ground state, using the muon mass in place of the electron mass.
  }
  \label{fig:gff}
\end{figure}

Lastly, for illustrative purposes, let us examine plots of numerical results
for the form factors.
The gravitational form factors are shown in Fig.~\ref{fig:gff}.
The hydrogen atom is an extremely mass-imbalanced system,
and to make the plots more instructive,
I use the muon mass $m_\mu \approx 105.66$~MeV in place of the physical
electron mass.
The proton dominates the form factor $A(\bd^2)$ because it contains most
of the atom's mass---even with the electron replaced by a muon.
By contrast, the proton makes comparatively small contributions to
$D(\bd^2)$ and $\bar{c}(\bd^2)$,
owing to the factors $m_p^{-3}$ and $m_p^{-1}$, respectively.
The Coulomb field makes the dominant contribution to $D(\bd^2)$.

%%%%%%%%%%%%%%%%%%%%%%%%%%%%%%%%%%%%%%%%%%%%%%%%%%%%%%%%%%%%%%%%%%%%%%%%%%%%%%%%

\section{Fourier transforms of form factors}
\label{sec:space}

Part of the motivation for studying gravitational form factors is that they
can be associated with spatial distributions.
In particular, given a physical state $|\psibar(t)\rangle$
for the atom's barycenter,
the expectation value of the stress tensor can be written:
\begin{align}
  \langle \psibar(t) | \hat{T}_a^{ij}(\bm{x}) | \psibar(t) \rangle
  =
  \int \frac{\d^3\bm{p}}{(2\pi)^3}
  \int \frac{\d^3\bm{p}'}{(2\pi)^3}
  \widetilde{\psibar}^*(\bm{p}',t)
  \widetilde{\psibar}(\bm{p},t)
  \langle \bm{p}' | \hat{T}_a^{ij}(0) | \bm{p} \rangle
  \e^{-\i\bd\cdot\bm{x}}
  \,,
\end{align}
where $\widetilde{\psibar}(\bm{p},t)$ is the barycentric wave packet in
momentum space.
Substituting in Eq.~(\ref{eqn:gff}),
and Fourier transforming the barycentric wave functions to configuration space gives:
\begin{align}
  \label{eqn:stress:exp}
  \langle \psibar(t) | \hat{T}_a^{ij}(\bm{x}) | \psibar(t) \rangle
  =
  \int \d^3 \bm{R} \,
  \left\{
    \left(
    -
    \psibar^*(\bm{R},t)
    \frac{\lrn^i \lrn^j}{4M}
    \psibar(\bm{R},t)
    \right)
    \bla_a(\bm{x} - \bm{R})
    +
    \blt_a^{ij}(\bm{x} - \bm{R})
    \psibar^*(\bm{R},t)
    \psibar(\bm{R},t)
    \right\}
  \,,
\end{align}
where I have defined the auxiliary quantities:
\begin{align}
  \label{eqn:bla}
  \bla_a(\bm{b})
  &=
  \int \frac{\d^3\bd}{(2\pi)^3}
  A_a(\bd^2)
  \e^{-\i\bd\cdot\bm{b}}
  \\
  \label{eqn:blt}
  \blt_a^{ij}(\bm{b})
  &=
  \int \frac{\d^3\bd}{(2\pi)^3}
  \left\{
    \frac{\varDelta^i \varDelta^j - \bd^2 \delta^{ij}}{4M}
    D_a(\bd^2)
    -
    M \delta^{ij} \bar{c}_a(\bd^2)
    \right\}
  \e^{-\i\bd\cdot\bm{b}}
  \,.
\end{align}
Here, $\bm{b} = \bm{x} - \bm{R}$ is a sort of impact parameter:
it's the displacement from the system's barycenter to the source point $\bm{x}$.
An additional quantity of interest is the Fourier transform of
$\bar{c}_a(\bd^2)$ by itself:
\begin{align}
  \label{eqn:blc}
  \blc_a(\bd^2)
  =
  \int \frac{\d^3\bd}{(2\pi)^2}
  \bar{c}_a(\bd^2)
  \e^{-\i\bd\cdot\bm{b}}
  \,.
\end{align}

The quantity $\blt^{ij}_a(\bm{b})$ is often interpreted as a sort of
intrinsic stress tensor,
and the piece involving $\bla_a(\bm{b})$ has been interpreted as
pressure exerted by dispersion of the barycentric wave packet.
(See Refs.~\cite{Freese:2021czn,Li:2022ldb,Freese:2022fat,Li:2024vgv} for instance.)
I will give a more complete appraisal of this interpretation in
Sec.~\ref{sec:bohm},
but this breakdown can be justified by examining very diffuse wave packets
with a mean momentum $\bm{p}_0$,
for which the $\bla_a(\bm{b})$ piece
looks like a dynamic pressure arising from motion of a
medium~\cite{binder1943fluid}.

It is often helpful to break
$\blt^{ij}_a(\bm{b})$
down into an isotropic piece and an anisotropic piece:
\begin{align}
  \blt^{ij}_a(\bm{b})
  &=
  \delta^{ij}
  \blp_a^{(\text{iso})}(\bm{b})
  +
  \left(
  \hat{b}^i \hat{b}^j - \frac{\delta^{ij}}{3}
  \right)
  \blp_a^{(\text{ani})}(\bm{b})
  \,.
\end{align}
Polyakov and Schweitzer~\cite{Polyakov:2018zvc}---as well as many other
authors---refer to the isotropic piece as the ``pressure''
and the anisotropic piece as the ``shear.''
The quantities can be written as one-dimensional integrals:
\begin{align}
  \notag
  \blp^{(\text{iso})}_a(\bm{b})
  &=
  -
  \frac{1}{2\pi^2}
  \int_0^\infty \d \varDelta \,
  \frac{\varDelta^4}{6M}
  D_a(\varDelta^2)
  j_0(\varDelta b)
  -
  M \blc_a(\bm{b})
  \\
  \label{eqn:sbes}
  \blp^{(\text{ani})}_a(\bm{b})
  &=
  -
  \frac{1}{2\pi^2}
  \int_0^\infty \d \varDelta \,
  \frac{\varDelta^4}{4M}
  D_a(\varDelta^2)
  j_2(\varDelta b)
  \,,
\end{align}
where $j_n(x)$ are the spherical Bessel functions.
Eq.~(\ref{eqn:sbes}) provides a practical pathway to numerically
calculate $\blt^{ij}_a(\bm{b})$
when deriving an analytic expression is difficult or impossible.

%%%%%%%%%%%%%%%%%%%%%%%%%%%%%%%%%%%%%%%%

\subsection{Particle contributions}

For particles ($a\in\{e,p\}$),
it is straightforward to obtain formulas for $\bla_a(\bm{b})$
and $\blt_a^{ij}(\bm{b})$;
the Fourier transforms of the relevant parts of
Eq.~(\ref{eqn:stress:chi})
simply produce delta functions upon integrating $\bd$.
These set $\bm{b} = \frac{\mu}{m_a} \bm{r}$;
the displacement of particle $a$ scales inversely
with respect to its mass at a fixed separation $\bm{r}$.
The resulting formulas are:
\begin{align}
  \label{eqn:A:b}
  \bla_a(\bm{b})
  &=
  \frac{m_a}{M}
  \left(\frac{m_a}{\mu}\right)^3
  \psirel^*\left(\frac{m_a b}{\mu}\right)
  \psirel\left(\frac{m_a b}{\mu}\right)
  \\
  \label{eqn:T:b}
  \blt^{ij}_a(\bm{b})
  &=
  -
  \frac{1}{4\mu}
  \psirel^*\left(\frac{m_a b}{\mu}\right)
  \lrn^i \lrn^j
  \psirel\left(\frac{m_a b}{\mu}\right)
  \,,
\end{align}
where the gradient is with respect to $\bm{b}$.
Using the explicit formula (\ref{eqn:wf}) for the ground state wave function gives:
\begin{align}
  \label{eqn:bla:ep}
  \bla_a(\bm{b})
  &=
  \frac{m_a}{M}
  \frac{(\alpha m_a)^3}{\pi}
  \e^{-2\alpha m_a b}
  \\
  \label{eqn:blt:ep}
  \blt^{ij}_a(\bm{b})
  &=
  \frac{(\alpha\mu)^3}{\pi}
  \frac{\alpha m_a}{2\mu}
  \frac{ \e^{-2\alpha m_a b} }{b}
  \big( \delta^{ij} - \hat{b}^i \hat{b}^j \big)
  \,.
\end{align}
Obtaining $\blc_a(\bm{b})$ requires working out the Fourier transform
of Eq.~(\ref{eqn:cbar:ep}).
The result is:
\begin{align}
  \label{eqn:cbar:b}
  \blc_a(\bm{b})
  &=
  \frac{-(\alpha m_a)^3}{\pi}
  \frac{\alpha \mu^2}{m_a^2 M b}
  E_2\Big( 2\alpha m_a b \Big)
  \,,
\end{align}
where $E_n(z) \equiv \int_1^\infty \d t \, t^{-n} \e^{-zt}$
is the generalized exponential integral function~\cite{NIST:DLMF}.

\begin{figure}
  \includegraphics[width=0.49\textwidth]{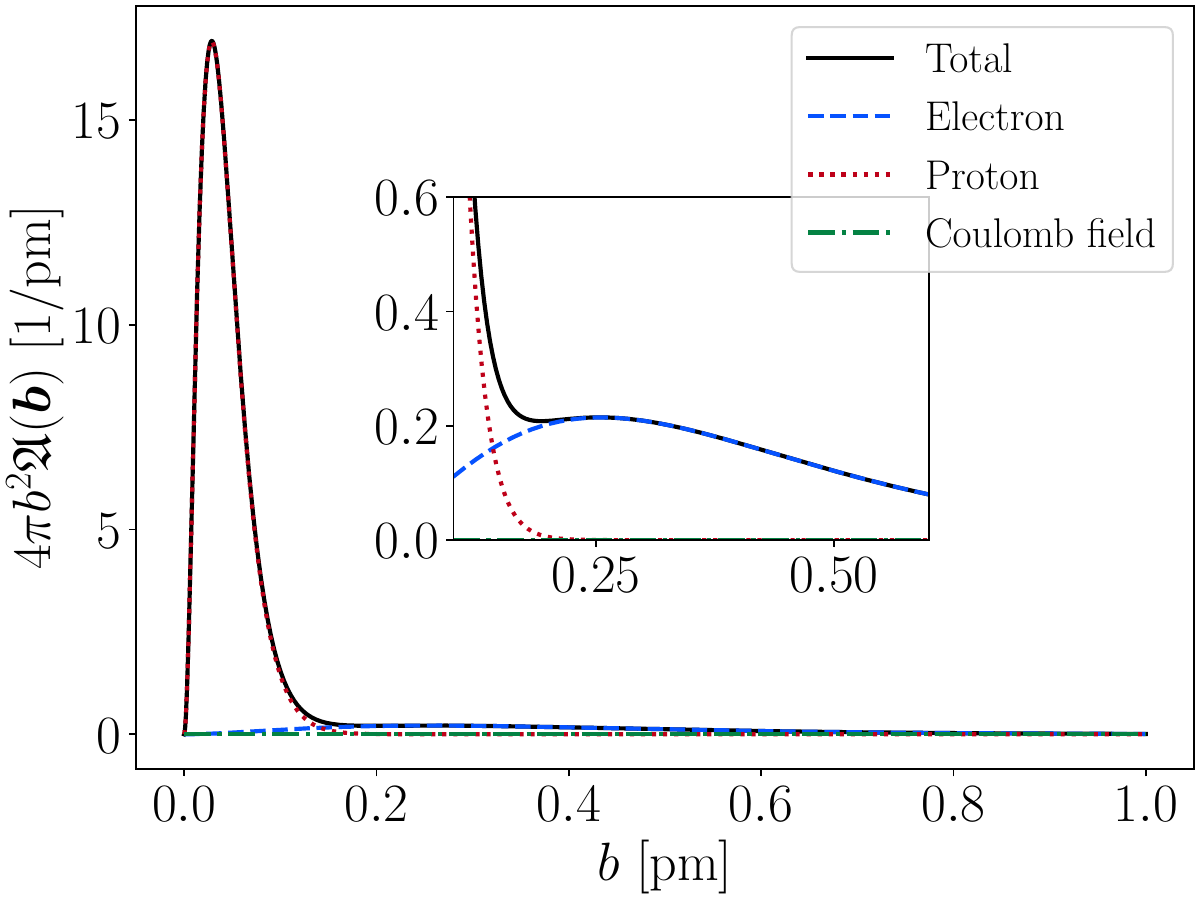}
  \includegraphics[width=0.49\textwidth]{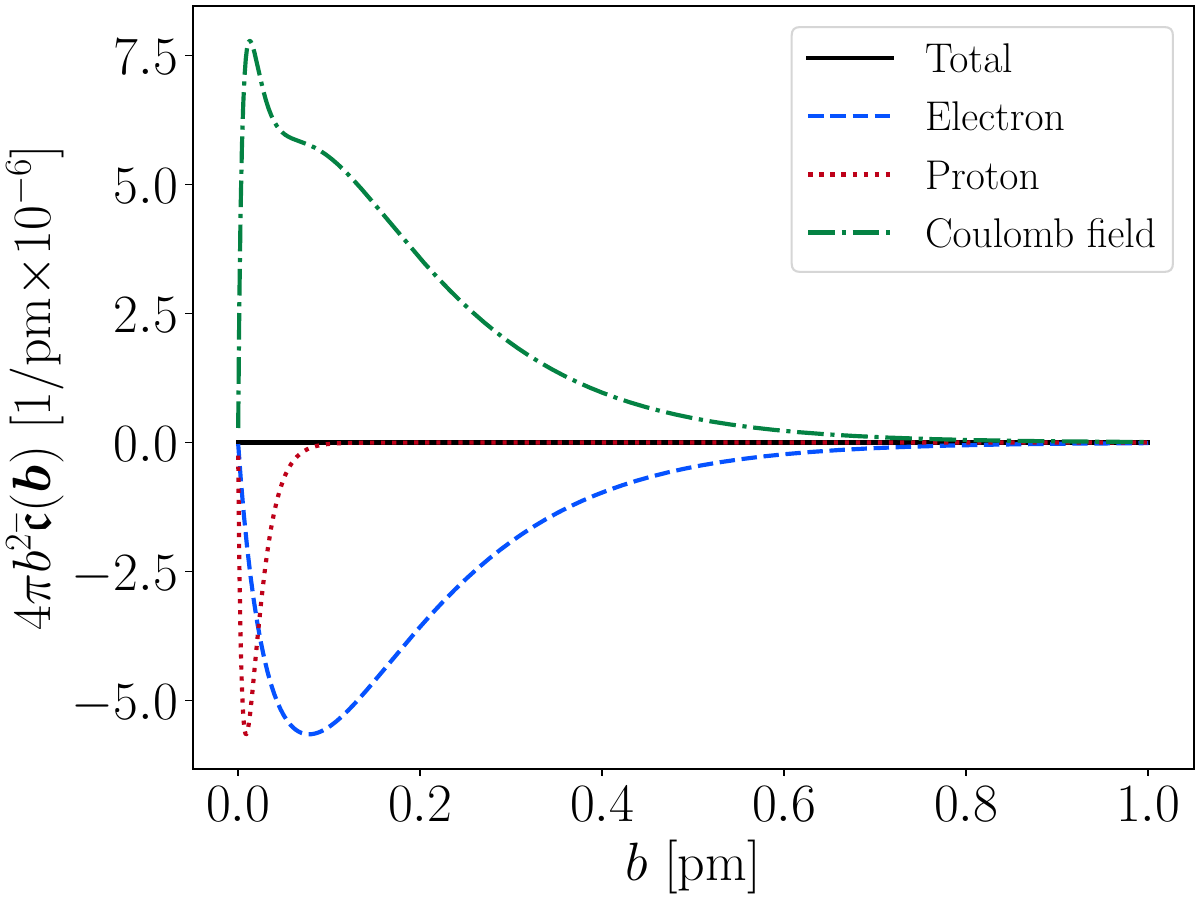}
  \caption{
    Fourier transforms of the form factors
    $A_a(\bd^2)$ (left panel)
    and $\bar{c}_a(\bd^2)$ (right panel).
    Contributions from the electron, proton and Coulomb field
    are present.
    The muon mass is used instead of the electron mass
    to make the mass imbalance less extreme.
    The plot of $\bla_A(\bm{b})$ contains an inlay
    to more easily see the diffuse muon cloud.
  }
  \label{fig:Ac:fourier}
\end{figure}

Numerical results for $\bla_a(\bm{b})$ and $\blc_a(\bm{b})$
are shown in Fig.~\ref{fig:Ac:fourier}.
Since $A_\gamma(\bd^2)=0$ and
$\bar{c}_\gamma(\bd^2) = - \bar{c}_e(\bd^2) - \bar{c}_p(\bd^2)$,
the Coulomb field contributions to the Fourier transforms
of these form factors are trivial, and are included in the
plots for completeness.
The quantity $\bla_a(\bm{b})$ can be interpreted as a mass fraction density,
since---via Eq.~(\ref{eqn:A:b})---it is simply the probability density
in terms of the physical impact parameter weighted by the mass fraction.
The left panel of Fig.~\ref{fig:Ac:fourier} then has a straightforward
interpretation: the majority of the mass is carried by the proton,
which is typically located close to the barycenter of the atom,
while the muon carries much less mass and is spread more diffusely
around the barycenter.

\begin{figure}
  \includegraphics[width=0.49\textwidth]{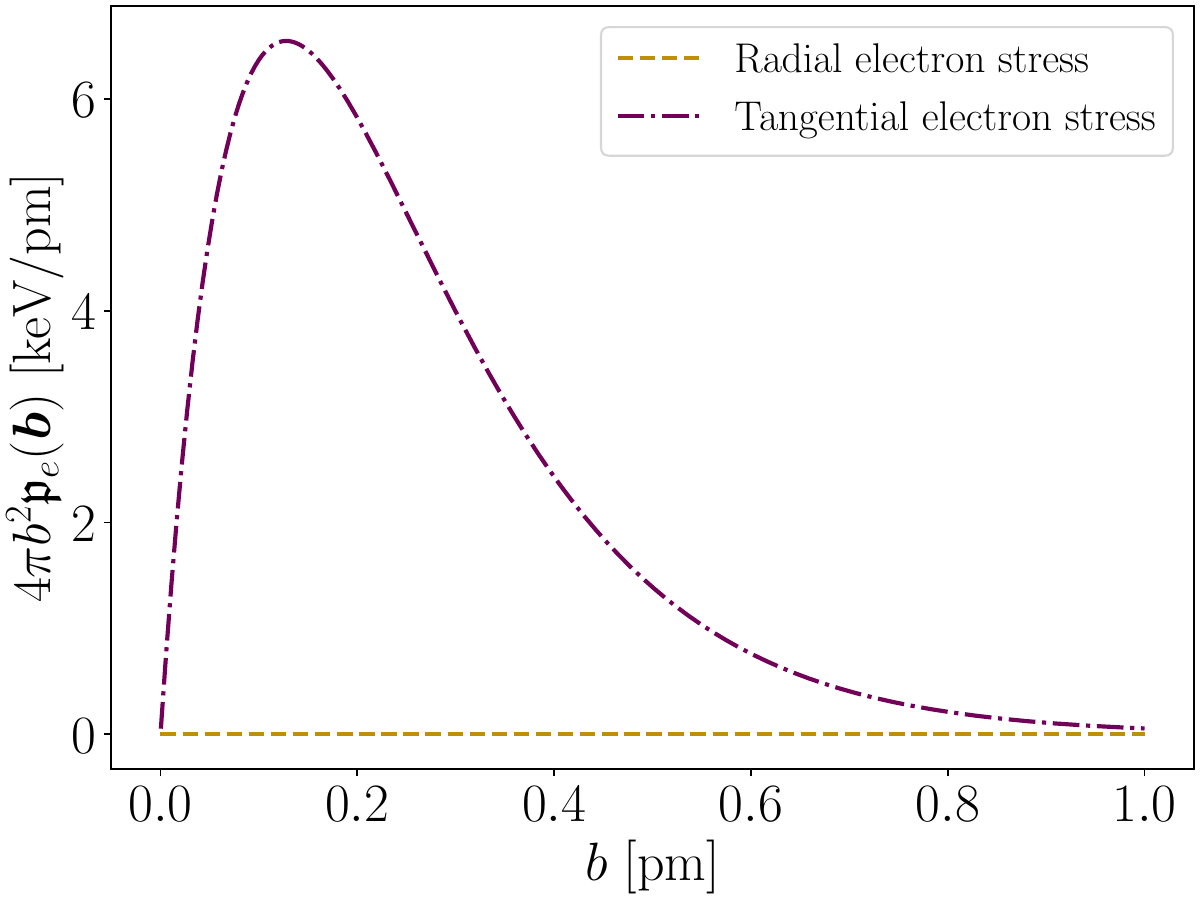}
  \includegraphics[width=0.49\textwidth]{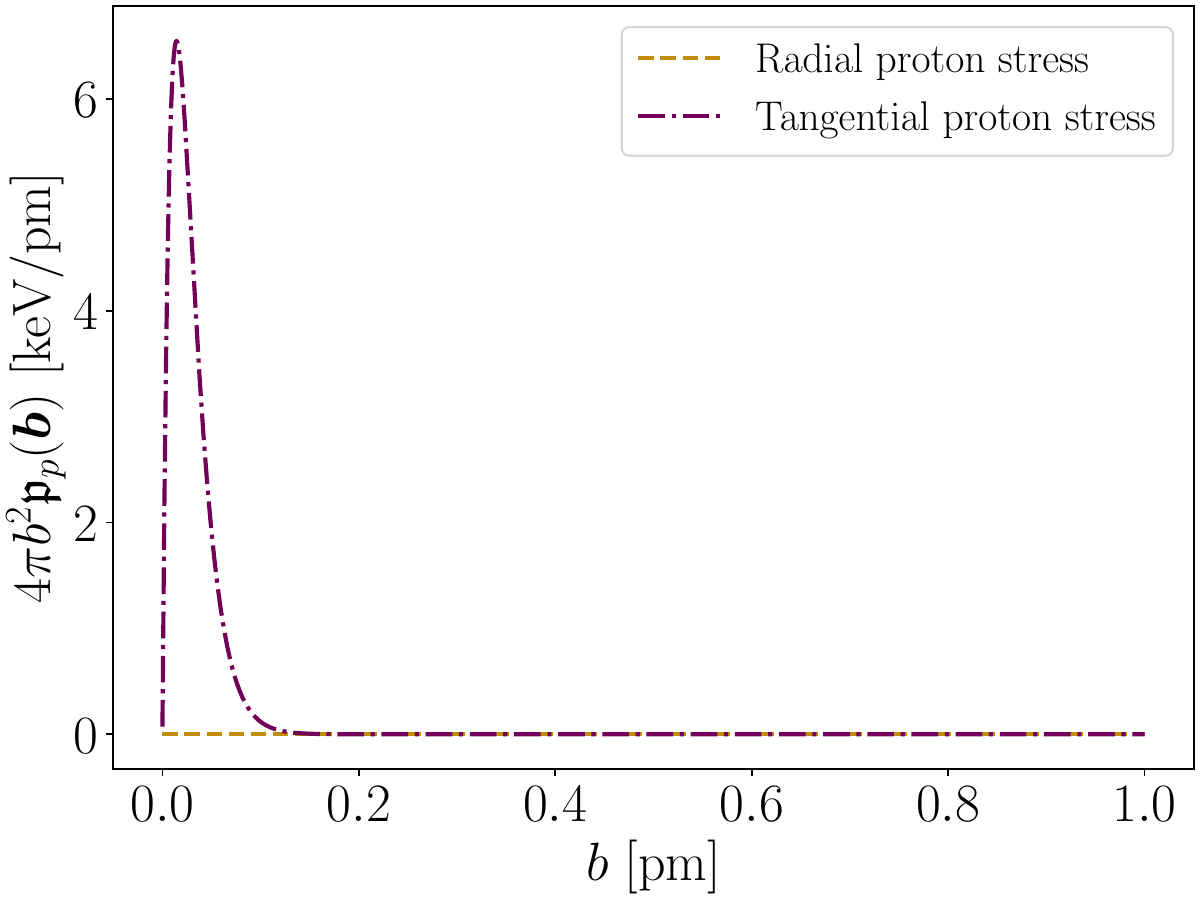}
  \caption{
    Radial and tangential components of
    $\blt^{ij}_{a}(\bm{b})$,
    for the electron (left panel) and the proton (right panel).
    The muon mass is used instead of the electron mass
    to make the mass imbalance less extreme.
  }
  \label{fig:prt}
\end{figure}

Numerical results for $\blt_{e,p}^{ij}(\bm{b})$
can be presented by projecting out components thereof.
As discussed by Polyakov and Schweitzer~\cite{Polyakov:2018zvc},
since $\blt_{e,p}^{ij}(\bm{b})$ is a $3\times3$ real symmetric matrix,
it can be diagonalized.
At any impact parameter $\bm{b}$,
the eigenvalues are given by
\begin{align}
  \notag
  \blp^{(r)}_a(\bm{b})
  & \equiv
  \hat{b}_i \hat{b}_j
  \blt_{e,p}^{ij}(\bm{b})
  \\
  \blp^{(t)}_a(\bm{b})
  & \equiv
  \hat{\theta}_i \hat{\theta}_j
  \blt_{e,p}^{ij}(\bm{b})
  =
  \hat{\phi}_i \hat{\phi}_j
  \blt_{e,p}^{ij}(\bm{b})
  \,,
\end{align}
which Polyakov and Schweitzer call the radial (or normal) pressure
and the tangential pressure.
I plot results for the electron and proton contributions to these
quantities in Fig.~\ref{fig:prt},
using the muon mass in place of the electron mass.
It is remarkable that the radial pressure is identically zero;
I discuss the implications of this later in Sec.~\ref{sec:potato}.
The tangential pressure is strictly positive,
and more tightly concentrated near the origin for the proton
than for the electron.

%%%%%%%%%%%%%%%%%%%%%%%%%%%%%%%%%%%%%%%%

\subsection{Coulomb field contributions}

\begin{figure}
  \includegraphics[width=0.49\textwidth]{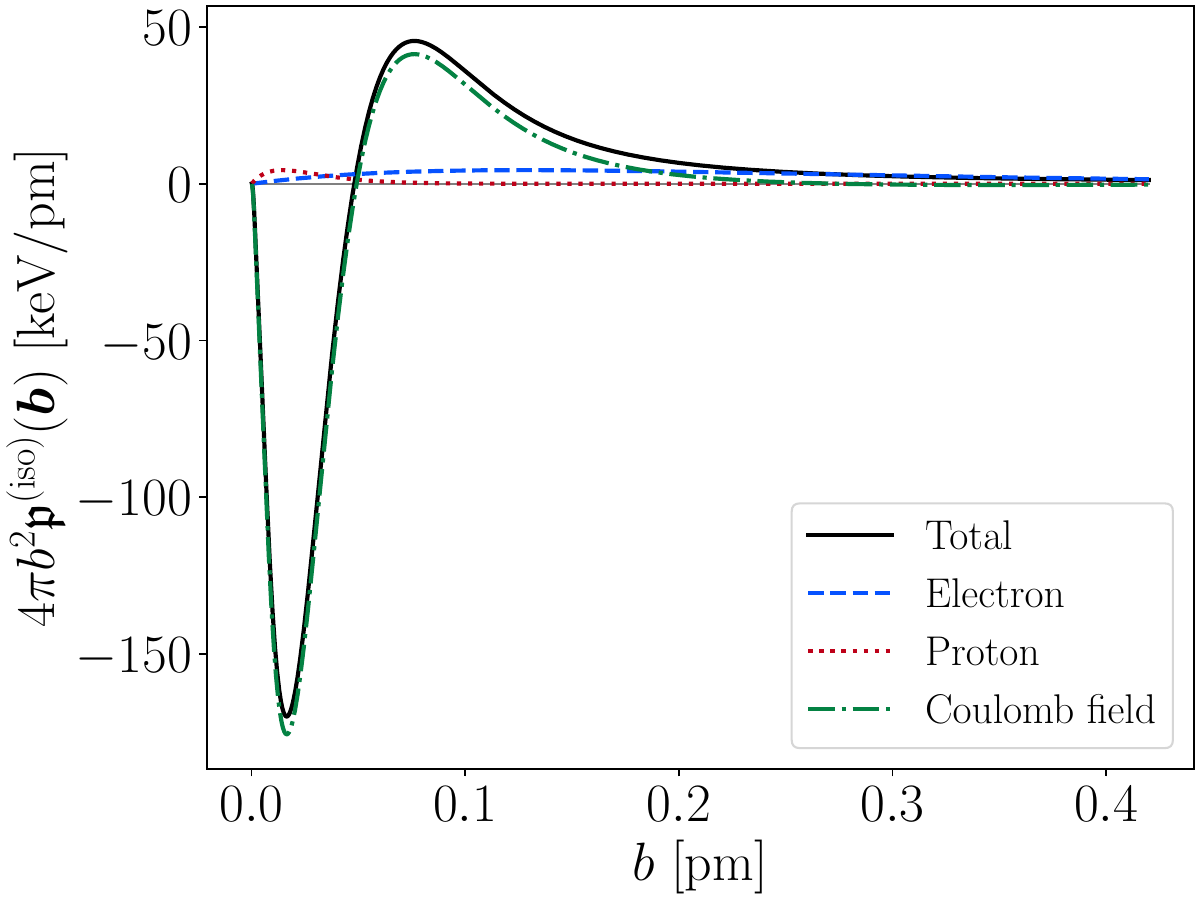}
  \includegraphics[width=0.49\textwidth]{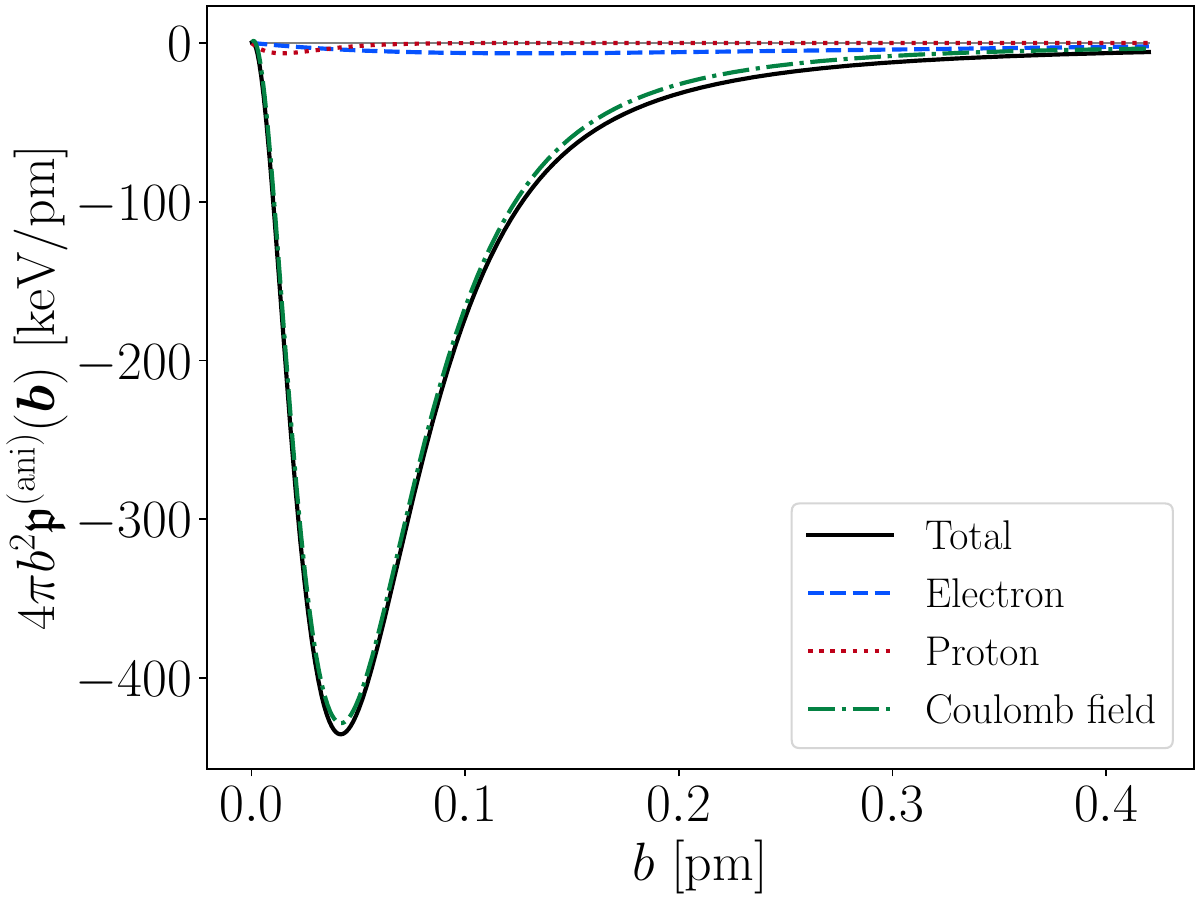}
  \caption{
    Isotropic (left panel) and anisotropic (right panel)
    pieces of the intrinsic stress tensor
    $\blt_a^{ij}(\bm{b})$,
    using the muon mass instead of the electron mass to make
    the mass imbalance less extreme.
    The electron (muon), proton and Coulomb field contributions
    are all included.
  }
  \label{fig:isoani}
\end{figure}

The Fourier transforms $\bla_\gamma(\bm{b})$ and $\blc_\gamma(\bm{b})$
are trivial:
\begin{align*}
  \bla_\gamma(\bm{b})
  &=
  0
  \\
  \blc_\gamma(\bm{b})
  &=
  -
  \sum_{a=e,p}
  \blc_a(\bm{b})
  \,.
\end{align*}
On the other hand, $\blt^{ij}_\gamma(\bm{b})$ is more complicated,
and I have opted to evaluate it numerically instead of analytically.
To this end, Eq.~(\ref{eqn:sbes}) can be used to obtain the isotropic
and anisotropic parts of $\blt^{ij}_\gamma(\bm{b})$.
Numerical results are shown in Fig.~\ref{fig:isoani},
along with the electron and proton contributions to the same quantities.
The Coulomb field contribution to the stress dominates over the particles.

%%%%%%%%%%%%%%%%%%%%%%%%%%%%%%%%%%%%%%%%

\subsection{Discussion}
\label{sec:potato}

So far, I have deferred interpretation of the Fourier transforms
of the gravitational form factors.
However, a peculiarity of the particle contributions to
the intrinsic stress tensor $\blt^{ij}_a(\bm{b})$ warrants attention.
The result, in Eq.~(\ref{eqn:blt:ep}) and Fig.~\ref{fig:prt},
is entirely tangential.
Any surface with a radial normal vector
$\hat{n} = \hat{b}$
has no momentum flux flowing through it.
This means that the electron and the proton carry and experience
no radial momentum flux---neither by particle motion nor by local stresses.

On the other hand,
a surface with a tangential normal vector
$\hat{n} = \cos(\eta) \hat{\theta} + \sin(\eta) \hat{\phi}$
has positive momentum flux flowing through it.
In general, this would be compatible with either particle motion
or with local stresses.
However, particle motion cannot be responsible for this momentum flux,
since that would imply circular motion,
and the state in question is an $\hat{\bm{L}}^2=0$ eigenstate.
The quantity $\blt^{ij}_{e,p}(\bm{b})$---if it has any meaning at all---must
therefore be provide local static forces experienced by the electron and proton.

Another peculiar implication of these results merits emphasis:
since the radial momentum flux is identically zero,
and---being a $\hat{\bm{L}}^2=0$ eigenstate---there is no tangential motion,
the electron and proton are stationary relative to the atom's barycenter.
As strange as this implication sounds,
it is actually a well-known result in the pilot wave
interpretation of quantum mechanics~\cite{Bohm:1951xw}.
It therefore seems appropriate to further investigate
how the stress tensor can be understood
in the pilot wave interpretation.

%%%%%%%%%%%%%%%%%%%%%%%%%%%%%%%%%%%%%%%%%%%%%%%%%%%%%%%%%%%%%%%%%%%%%%%%%%%%%%%%

\section{Pilot wave interpretation of the quantum stress tensor}
\label{sec:bohm}

The pilot wave interpretation is a formulation of quantum mechanics
in which particles have sharply-defined positions
and are guided in their motion by the wave function.
It was first discovered by de Broglie in 1927~\cite{db:pilot},
and rediscovered by Bohm in 1952~\cite{Bohm:1951xw,Bohm:1951xx};
it is thus sometimes called de Broglie-Bohm theory
or Bohmian mechanics\footnote{
  Bohm's own preferred moniker was the ontological interpretation~\cite{Bohm:2006und}.
}.
Excellent expositions of the subject can be found in
Refs.~\cite{Bohm:2006und,holland1995quantum}.
Since the pilot wave interpretation may be unfamiliar to many readers,
I will present the basic formulation before
applying it to the matter of the stress tensor.

In the pilot wave interpretation,
a complete description of
a non-relativistic quantum mechanical system consists of both a
wave function and a trajectory.
The wave function
$\Psi(\bm{q},t)$
is, as usual, a function of both the system's configuration
$\bm{q}$ and time,
while the trajectory $\bm{q}(t)$ is a function from time to configuration
space\footnote{
  For a system of $N$ spin-zero particles the configuration
  $\bm{q} = (\bm{q}_1, \bm{q}_2, \ldots, \bm{q}_N)$
  is simply a list of the particles' positions,
  and the configuration space is $\mathbb{R}^{3N}$.
}.
The pilot wave interpretation is in effect a ``hidden variable''
formulation of quantum mechanics,
with the exact trajectories as the hidden variables.

The wave function obeys the usual quantum mechanical wave equation,
which for $N$ spin-zero particles is the Schr\"{o}dinger equation\footnote{
  In contrast to the rest of the paper,
  I am keeping $\hslash\neq1$ in this section.
}:
\begin{align}
  \label{eqn:schrodinger}
  \i \hslash
  \frac{\partial \Psi(\bm{q},t)}{\partial t}
  =
  -\sum_{n=1}^N
  \frac{\hslash^2}{2m_n} \bm{\nabla}_n^2
  \Psi(\bm{q},t)
  +
  V(\bm{q},t)
  \Psi(\bm{q},t)
  \,.
\end{align}
The trajectory obeys a velocity law, which depends on the wave function:
\begin{align}
  \frac{\d \bm{q}(t)}{\d t}
  =
  f\big(\Psi(\bm{q},t)\big)
  \,.
\end{align}
Thus, given an initial condition $\bm{q}(t_0) = \bm{q}_0$ and the wave function
$\Psi(\bm{q},t)$, the trajectory is fully determined.
Our inability to predict the configuration of a system---and thus the outcome
of a measurement---is merely due to our
ignorance of its initial conditions,
to which we must assign a probability distribution.

The exact form of the velocity law is derived from a polar decomposition
of the wave function~\cite{Bohm:1951xw,Bohm:1951xx}:
\begin{align}
  \label{eqn:polar}
  \Psi(\bm{q},t)
  =
  \mathscr{R}(\bm{q},t)
  \e^{\i \mathscr{S}(\bm{q},t) / \hslash}
  \,.
\end{align}
Plugging this into the Schr\"{o}dinger equation (\ref{eqn:schrodinger})
gives two coupled equations
for the real-valued functions $\mathscr{R}$ and $\mathscr{S}$.
The first:
\begin{align}
  \frac{\d \mathscr{S}(\bm{q},t)}{\d t}
  +
  \sum_{n=1}^N
  \frac{(\bm{\nabla}_n\mathscr{S}(\bm{q},t))^2}{2m_n}
  +
  V(\bm{q},t)
  -
  \sum_{n=1}^N
  \frac{\hslash^2}{2m_n}
  \frac{\bm{\nabla}_n^2\mathscr{R}(\bm{q},t)}{\mathscr{R}(\bm{q},t)}
  =
  0
  \,,
\end{align}
takes the form of the Hamilton-Jacobi equation,
with $\mathscr{S}$ playing the role of Hamilton's principal function.
Taking this form at face value suggests the system experiences an
effective potential $V(\bm{q},t) + Q(\bm{q},t)$, where
\begin{align}
  \label{eqn:qpot}
  Q(\bm{q},t)
  =
  -
  \sum_{n=1}^N
  \frac{\hslash^2}{2m_n}
  \frac{\bm{\nabla}_n^2\mathscr{R}(\bm{q},t)}{\mathscr{R}(\bm{q},t)}
\end{align}
is called the quantum potential~\cite{Bohm:1951xw,Bohm:1951xx}.
It also suggests the particles making up the system have momenta:
\begin{align}
  \bm{p}_n^{\bohm}(\bm{q},t)
  &=
  \bm{\nabla}_n
  \mathscr{S}(\bm{q},t)
\end{align}
when the system is in the configuration $\bm{q}$,
where I have followed Flack and Hiley~\cite{Flack:2015cam}
in labeling this as the ``Bohm momentum,''
to contrast it with the canonical momentum.
Taking $\bm{p}_n^{\bohm}(\bm{q},t) = m_n \dot{\bm{q}}_n(t)$
provides the desired velocity law:
\begin{align}
  \label{eqn:velocity}
  \frac{\d \bm{q}_n(t)}{\d t}
  =
  \frac{\bm{\nabla}_n \mathscr{S}(\bm{q},t)}{m_n}
  \,.
\end{align}
In general, this law is extremely non-local:
the velocity of the $n$th particle depends not only on its position,
but the configuration of the system as a whole---including the
positions of all other particles.
This non-locality is a necessity;
Bell proved that any hidden variable formulation
of quantum mechanics must have these kinds of non-locality
to reproduce the statistical predictions
of quantum mechanics~\cite{Bell:1964kc,Bell:2004gpx}.

The second equation takes the form of a continuity equation:
\begin{align}
  \frac{\d \mathscr{R}^2(\bm{q},t)}{\d t}
  +
  \sum_{n=1}^N
  \bm{\nabla}_n \cdot
  \left[
    \frac{\bm{\nabla}_n \mathscr{S}(\bm{q},t)}{m_n}
    \mathscr{R}^2(\bm{q},t)
    \right]
  =
  0
  \,.
\end{align}
Given the velocity law (\ref{eqn:velocity}),
this can be rewritten:
\begin{align}
  \frac{\d \mathscr{R}^2(\bm{q},t)}{\d t}
  +
  \bm{\nabla} \cdot
  \left[
    \dot{\bm{q}}(t)
    \mathscr{R}^2(\bm{q},t)
    \right]
  =
  0
  \,.
\end{align}
This tells us that if we use Born's rule
($\mathscr{P}(\bm{q},t) = \mathscr{R}^2(\bm{q},t) = |\Psi(\bm{q},t)|^2$)
to assign a probability distribution over configuration space at any time,
Born's rule provides the correct probability distribution at all other times.

In effect, the pilot wave interpretation supplements vanilla quantum mechanics
with exact particle positions and trajectories.
The primary appeal of this interpretation in the big picture is that it
provides a clear answer to the measurement problem,
as most poignantly expressed by J.S.\ Bell~\cite{Bell:1966dtm,Bell:1989qe,Bell:2004gpx}.
The small-picture motivation relevant to this work, however,
is that the pilot wave interpretation allows questions about
the electron's and proton's positions and velocities,
and the forces acting on them,
to be clearly asked and answered.
It thus becomes possible to address the matter
of whether the stress tensor quantifies particle motion or local static forces.

%%%%%%%%%%%%%%%%%%%%%%%%%%%%%%%%%%%%%%%%

\subsection{Stress tensor for a free particle}

For a single free particle, the expectation value of the stress tensor is:
\begin{align}
  \langle \Psi(t) | \hat{T}^{ij}(\bm{x}) | \Psi(t) \rangle
  =
  -
  \frac{\hslash^2}{4m}
  \Psi^*(\bm{x},t)
  \lrn^i \lrn^j
  \Psi(\bm{x},t)
  \,.
\end{align}
In terms of the polar decomposition (\ref{eqn:polar}),
this can be rewritten:
\begin{align}
  \langle \Psi(t) | \hat{T}^{ij}(\bm{x}) | \Psi(t) \rangle
  =
  \mathscr{R}^2
  \left\{
    \frac{ (\nabla^i\mathscr{S}) (\nabla^j\mathscr{S}) }{m}
    +
    \frac{\hslash^2}{2m\mathscr{R}^2}
    \Big(
    (\nabla^i\mathscr{R}) (\nabla^j\mathscr{R})
    -
    \mathscr{R} (\nabla^i\nabla^j\mathscr{R})
    \Big)
    \right\}
  \,,
\end{align}
where I have suppressed the $(\bm{x},t)$ dependence of $\mathscr{R}$
and $\mathscr{S}$ to make the formula more compact.
Given the velocity law (\ref{eqn:velocity}), this can again be rewritten:
\begin{align}
  \langle \Psi(t) | \hat{T}^{ij}(\bm{x}) | \Psi(t) \rangle
  =
  \mathscr{R}^2
  \left\{
    \frac{p^i_{\bohm}p^j_{\bohm}}{m}
    +
    \frac{\hslash^2}{2m\mathscr{R}^2}
    \Big(
    (\nabla^i\mathscr{R}) (\nabla^j\mathscr{R})
    -
    \mathscr{R} (\nabla^i\nabla^j\mathscr{R})
    \Big)
    \right\}
  \,.
\end{align}
The stress tensor of the free particle is a sum of the classical stress tensor
(weighted by the probability $\mathscr{R}^2$, since this is an expectation value)
and a quantum stress tensor:
\begin{align}
  T_{Q}^{ij}(\bm{x},t)
  =
  \frac{\hslash^2}{2m}
  \Big(
  (\nabla^i\mathscr{R}) (\nabla^j\mathscr{R})
  -
  \mathscr{R} (\nabla^i\nabla^j\mathscr{R})
  \Big)
  \,.
\end{align}
The quantum stress tensor is intimately connected to Bohm's quantum potential
(\ref{eqn:qpot}), and in effect quantifies the spatial density of the forces
contributing to the quantum potential.
Taking minus its divergence gives:
\begin{align}
  -
  \nabla_i
  T_{Q}^{ij}(\bm{x},t)
  =
  \mathscr{R}^2
  \frac{\hslash^2}{2m}
  \nabla^j
  \frac{\bm{\nabla}^2\mathscr{R}}{\mathscr{R}}
  =
  -
  \mathscr{R}^2
  \nabla^j
  Q(\bm{x},t)
  \,,
\end{align}
which is exactly the force associated with the quantum potential
(again weighted by the probability density).
In accordance with the continuity equation (\ref{eqn:flux}),
the divergence
$ - \nabla_i T_{Q}^{ij}(\bm{x},t) $
constitutes part of the momentum flux density,
and its relation to the quantum potential tells us
that it arises from local forces.
These forces can be thought of as forces exerted on the particle
by the wave function.
Since the particle is inseparable from the wave function---the two forming
a holistic whole~\cite{Bohm:2006und},
with the behavior of the former determined fully by the latter---these
are in effect internal rather than external forces.

%%%%%%%%%%%%%%%%%%%%%%%%%%%%%%%%%%%%%%%%

\subsection{Two-body systems and barycentric coordinates}

As usual, it is convenient to factorize the wave function for two-body
bound systems into barycentric and internal wave functions,
$\psibar(\bm{R},t)$ and $\psirel(\bm{r},t)$ respectively.
Because they factorize,
the polar decomposition (\ref{eqn:polar}) can be done separately for each:
\begin{align}
  \notag
  \psibar(\bm{R},t)
  &=
  \mathscr{R}(\bm{R},t)
  \e^{\i\mathscr{S}(\bm{R},t)/\hslash}
  \\
  \psirel(\bm{r},t)
  &=
  \mathcal{R}(\bm{r},t)
  \e^{\i\mathcal{S}(\bm{r},t)/\hslash}
  \,,
\end{align}
where I have used script symbols for the barycentric wave packet
and calligraphic symbols for the internal wave function.
Now, if the polar decomposition of the barycentric wave packet
is applied to the stress tensor breakdown (\ref{eqn:stress:exp}),
then:
\begin{multline}
  \label{eqn:bohm:disp}
  \langle \psibar(t) | \hat{T}_a^{ij}(\bm{x}) | \psibar(t) \rangle
  =
  \int \d^3 \bm{R} \,
  \mathscr{R}^2(\bm{R},t)
  \Bigg\{
    \frac{p^i_{\bohm}(\bm{R},t)p^j_{\bohm}(\bm{R},t)}{M}
    \bla_a(\bm{x}-\bm{R})
    +
    \blt_a^{ij}(\bm{x} - \bm{R})
    \\
    +
    \frac{\hslash^2}{2M\mathscr{R}^2(\bm{R},t)}
    \Big[
    \big(\nabla^i\mathscr{R}(\bm{R},t)\big) \big(\nabla^j\mathscr{R}(\bm{R},t)\big)
    -
    \mathscr{R}(\bm{R},t) \big(\nabla^i\nabla^j\mathscr{R}(\bm{R},t)\big)
    \Big]
    \bla_a(\bm{x}-\bm{R})
    \Bigg\}
  \,.
\end{multline}
Several authors~\cite{Freese:2021czn,Li:2022ldb,Freese:2022fat,Li:2024vgv}
have previously claimed that the part of the stress tensor containing
$\bla_a(\bm{b})$ encodes stresses from barycentric motion and wave packet
dispersion, and
Eq.~(\ref{eqn:bohm:disp}) validates these claims.
The $\bla_a(\bm{b})$ term here is cleanly broken into two pieces.
The first piece involves the Bohm momentum $\bm{p}_{\bohm}$,
and gives the classical expression for momentum flux density of a moving particle.
The second piece---on the second line of Eq.~(\ref{eqn:bohm:disp})---contains
the quantum stress tensor and describes forces exerted by
the \emph{barycentric} wave packet on the constituent $a$.
Since these forces are exerted by the barycentric wave packet,
they are separate from internal forces,
and the standard procedure of excluding $A_a(\bd^2)$
from intrinsic stress densities is fully justified.

The intrinsic stress tensor can also be broken down using
polar decomposition.
For the particle contributions,
plugging the polar decomposition into Eq.~(\ref{eqn:T:b})
(with appropriate factors of $\hslash$ restored) gives:
\begin{multline}
  \label{eqn:blt:bohm}
  \blt^{ij}_a(\bm{b})
  =
  \left(\frac{m_a}{\mu}\right)^2
  \mathcal{R}^2(\bm{r},t)
  \Bigg\{
    \frac{k_{\bohm}^i(\bm{r},t) k_{\bohm}^j(\bm{r},t)}{\mu}
    \\
    +
    \frac{\hslash^2}{2\mu\mathcal{R}^2(\bm{r},t)}
    \Big[
    \big(\nabla^i\mathcal{R}(\bm{r},t)\big) \big(\nabla^j\mathcal{R}(\bm{r},t)\big)
    -
    \mathcal{R}(\bm{r},t) \big(\nabla^i\nabla^j\mathcal{R}(\bm{r},t)\big)
    \Big]
    \Bigg\}
  \Bigg|_{\bm{r} = m_a\bm{b}/\mu}
  \,,
\end{multline}
where $\bm{k}_{\bohm}(\bm{r},t) = \bm{\nabla} \mathcal{S}(\bm{r},t)$
is the relative Bohm momentum,
and the gradients are all with respect to $\bm{r}$ rather than $\bm{b}$
(with the chain rule having been applied as appropriate).
Just as in the free case,
this breaks down into a classical expression involving particle motion
and a new quantum stress tensor:
\begin{align}
  \blq^{ij}_{a}(\bm{b})
  \equiv
  \left(\frac{m_a}{\mu}\right)^2
  \frac{\hslash^2}{2\mu}
  \bigg[
    \big(\nabla^i\mathcal{R}(\bm{r},t)\big) \big(\nabla^j\mathcal{R}(\bm{r},t)\big)
    -
    \mathcal{R}(\bm{r},t) \big(\nabla^i\nabla^j\mathcal{R}(\bm{r},t)\big)
    \bigg]
  \,,
\end{align}
and just as in the free case, the quantum stress tensor can be related to
the quantum potential:
\begin{align}
  -
  \nabla_i
  \blq^{ij}_{a}(\bm{b})
  =
  \left(\frac{m_a}{\mu}\right)^3
  \mathcal{R}^2
  \frac{\hslash^2}{2m}
  \nabla^j
  \frac{\bm{\nabla}^2\mathcal{R}}{\mathcal{R}}
  =
  -
  \left(\frac{m_a}{\mu}\right)^3
  \mathcal{R}^2
  \nabla^j
  Q_{\text{rel.}}(\bm{r},t)
  \,.
\end{align}
Thus, in general, the intrinsic stress tensor
$\blt^{ij}_a(\bm{b})$ contains contributions from both particle motion
and from local stresses,
the latter quantifying forces exerted by the internal wave function
$\psirel(\bm{r},t)$ on the particle.

In the special case of two-body systems in $m_l=0$ states---including
even $l\neq0$ states---the phase of $\psirel(\bm{r},t)$ is independent
of position, since the position dependence of the phase comes
entirely from $\e^{\i m_l \phi}$.
Thus, $\bm{k}_{\bohm} = 0$ for these states,
and the intrinsic stress tensor can be entirely attributed to forces
exerted by the wave function on the particle.
This is the case for the hydrogen atom's ground state.

%%%%%%%%%%%%%%%%%%%%%%%%%%%%%%%%%%%%%%%%

\subsection{$\bar{c}~$ and Cauchy's first law of motion}

For continuum systems,
Cauchy's first law of
motion~\cite{chatterjee1999mathematical,irgens2008continuum}
(sometimes called Cauchy's equation of
motion~\cite{chadwick1999continuum,chen2006meshless,irgens2008continuum,nayak2022continuum})
says:
\begin{align}
  \label{eqn:cauchy}
  \frac{\d}{\d t}
  \Big[
    T^{0j}(\bm{x},t)
    \Big]
  +
  \nabla_i
  T^{ij}(\bm{x},t)
  =
  \mathscr{P}(\bm{x},t)
  F^j_{\text{ext}}(\bm{x},t)
  \,,
\end{align}
where $T^{0j}(\bm{b},t)$ is the momentum density,
$\mathscr{P}(\bm{b},t)$ is the number density
(the probability density in this case),
and $F^j_{\text{ext}}$ is any external force
(sometimes called a body force~\cite{chatterjee1999mathematical,chadwick1999continuum,irgens2008continuum,nayak2022continuum}).
It is in essence a continuum version of Newton's second law of motion.
This law can also be thought of as a generalization of the
continuity equation (\ref{eqn:cont}) for open systems.

Let us apply Cauchy's first law to the
internal stress tensor (\ref{eqn:blt:bohm}) experienced by the particles,
specifically for stationary states.
The momentum density for stationary states is constant in time, so:
\begin{align}
  \nabla_i
  \blt^{ij}_a(\bm{b})
  =
  \left(\frac{m_a}{\mu}\right)^3
  \mathcal{R}^2(\bm{r},t)
  F_{\text{ext}}^j(\bm{r},t)
  \bigg|_{\bm{r}=m_a\bm{m}/\mu}
  \,,
\end{align}
where the factor $(m_a/\mu)^3$ is a Jacobian that normalizes the probability
when integrated over the physical impact parameter $\bm{b}$.
Comparing to Eqs.~(\ref{eqn:blt}) and (\ref{eqn:blc}), this entails:
\begin{align}
  \label{eqn:cbar:force}
  -
  M
  \nabla^j
  \blc_a(\bm{b})
  =
  \left(\frac{m_a}{\mu}\right)^3
  \left|
  \psirel
  \left(\frac{m_a\bm{b}}{\mu}\right)
  \right|^2
  F_{\text{ext}}^j
  \left(\frac{m_a\bm{b}}{\mu}\right)
  \,.
\end{align}
The term involving $D_a(\bd^2)$ in Eq.~(\ref{eqn:blt}) dropped out
because its divergence is zero.

In the present context,
the ``external'' force is the electrostatic force.
We are after all examining stresses experienced by the electron
and the proton, which include particle motion and forces exerted
by the pilot wave---effectively, forces ``internal'' to the particles.
The only additional force acting on the particles is the electrostatic force
between them, which therefore is the external force.
This can be verified directly:
taking the gradient of $\blc_a(\bm{b})$ in Eq.~(\ref{eqn:cbar:b}),
using the wave function of Eq.~(\ref{eqn:wf}),
and plugging these into Eq.~(\ref{eqn:cbar:force}) gives:
\begin{align}
  F_{\text{ext}}^j
  \left(\frac{m_a\bm{b}}{\mu}\right)
  =
  -
  \frac{\alpha\hat{b}}{(m_ab/\mu)^2}
  \,,
\end{align}
which is exactly the Coulomb force law for a relative separation
$\bm{r} = m_n\bm{b}/\mu$.
In effect what this demonstrates is that if we had empirical access
to the probability density and to the form factor $\bar{c}_a(\bd^2)$,
we could reconstruct the force law binding the system,
even if we were previously ignorant of it.

%%%%%%%%%%%%%%%%%%%%%%%%%%%%%%%%%%%%%%%%

\subsection{Coulomb field stress}

The stress tensor of the Coulomb field,
$\blt_\gamma^{ij}(\bm{b})$,
is less remarkable from the pilot wave perspective.
The Coulomb field was treated essentially classically in this calculation,
albeit with the locations of its
sources promoted to operators.
In the pilot wave interpretation,
the electron and proton have definite positions
that we are merely ignorant of, and accordingly the electric field
has a definite, dipole-like configuration as well.
The stress tensor $\blt_\gamma^{ij}(\bm{b})$
is simply equal to the ensemble average of the classical stress tensor
$-E^i E^j + \frac{1}{2} \delta^{ij} \bm{E}^2$
evaluated for these definite configurations.

\begin{figure}
  \includegraphics[width=0.49\textwidth]{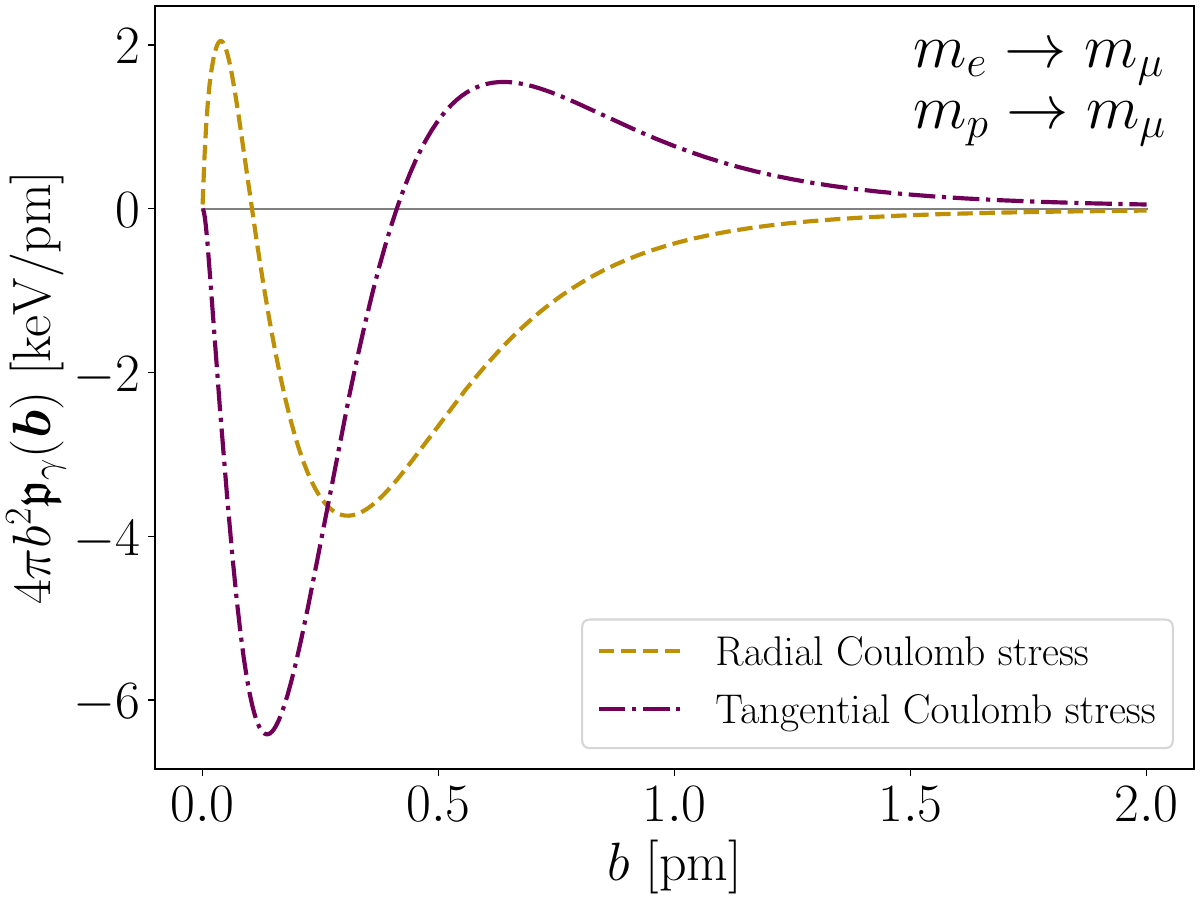}
  \includegraphics[width=0.49\textwidth]{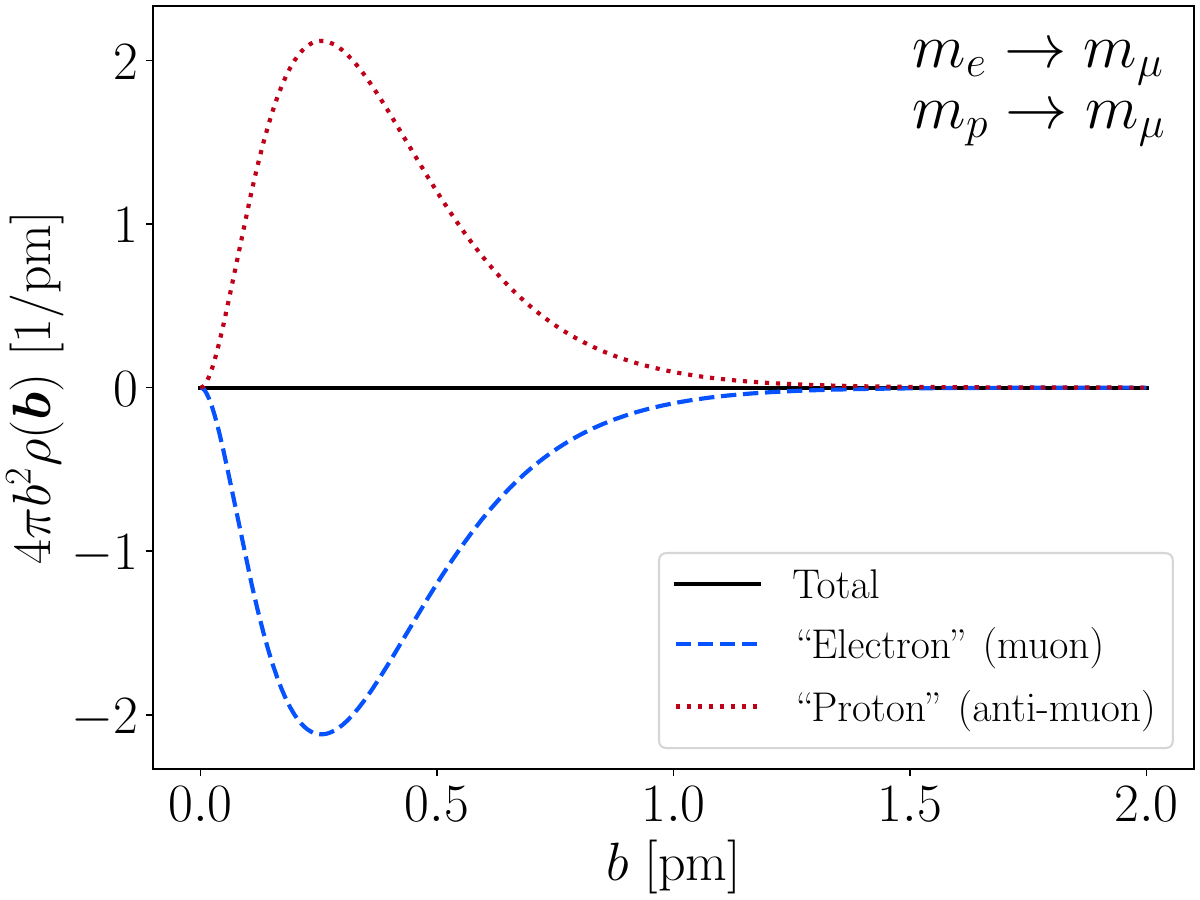}
  \caption{
    Plots of the radial and tangential components of the Coulomb
    stress tensor (left panel)
    and the charge density (right panel)
    in a mass-balanced variation of the hydrogen atom.
    The muon mass has been used in place of both the electron
    and proton masses in these plots.
  }
  \label{fig:balance}
\end{figure}

There are a few important conceptual points to bring up
in this context, however.
The first is that the hydrogen atom
should not be thought of as a classical charge distribution with charge density
\begin{align*}
  \langle \rho(\bm{b}) \rangle
  =
  e\left\{
    \left(\frac{m_p}{\mu}\right)^2
    \left|\psirel\left(\frac{m_p \bm{b}}{\mu}\right)\right|^2
    -
    \left(\frac{m_e}{\mu}\right)^2
    \left|\psirel\left(\frac{m_e \bm{b}}{\mu}\right)\right|^2
    \right\}
  \,.
\end{align*}
Although this is the expectation value of the charge density,
this expectation value is the ensemble average of the charge distributions
in actual, definite configurations.
In particular, it would be mistaken to assume that the electric field is given
by $\bm{\nabla}\cdot\bm{E} = \langle\rho(\bm{b})\rangle$.
In fact, only the \emph{expectation value} of the electric field is given by
this equation:
\begin{align}
  \bm{\nabla}\cdot
  \langle \bm{E}(\bm{b}) \rangle
  =
  \langle \rho(\bm{b}) \rangle
  \,.
\end{align}
On the other hand, expectation values of products are not
products of expectation values:
\begin{align}
  \left\langle
  -E^i(\bm{b}) E^j(\bm{b})
  +
  \frac{1}{2} \delta^{ij} \bm{E}^2(\bm{b})
  \right\rangle
  \neq
  -
  \langle E^i(\bm{b})\rangle
  \langle E^j(\bm{b})\rangle
  +
  \frac{1}{2} \delta^{ij}
  \langle \bm{E}(\bm{b}) \rangle^2
  \,,
\end{align}
and the Coulomb field contribution to
the stress tensor of the mass-balanced system (unlike the charge density)
is not zero.
This is demonstrated numerically in Fig.~\ref{fig:balance}.

To be sure, this is a basic property of expectation values,
and the results in Fig.~\ref{fig:balance} do not
rely on the pilot wave interpretation.
However, the claim of the pilot wave formulation
that every individual atom is different---and
that our spatial densities
(such as the charge density and Coulomb stress tensor
in Fig.~\ref{fig:balance})
are ensemble averages---fits very snugly
into these results.

The second conceptual point to raise is that since the Coulomb field is
treated classically,
its stress tensor can also be interpreted classically.
The Coulomb stress tensor $\blt^{ij}_\gamma(\bm{b})$,
with its contributions from $D_\gamma(\bd^2)$ and $\bar{c}_\gamma(\bd^2)$,
encodes actual stresses inside the hydrogen atom just as much
as a classical electrostatic field carries stresses.
Since I previously concluded that the particle stress tensors
$\blt_{e,p}^{ij}(\bm{b})$ can be interpreted as actual stresses,
it follows that the stress tensor as a whole can be interpreted as
encoding literal stresses for the hydrogen atom's ground state.

%%%%%%%%%%%%%%%%%%%%%%%%%%%%%%%%%%%%%%%%%%%%%%%%%%%%%%%%%%%%%%%%%%%%%%%%%%%%%%%%

\section{Discussion and outlook}
\label{sec:end}

In this work, I calculated gravitational form factors
and their Fourier transforms for the hydrogen atom in its ground state,
and then used the pilot wave interpretation of quantum mechanics to
interpret the results.
The impact parameter expressions from the Fourier transforms
are related to expectation values of the stress tensor through
Eq.~(\ref{eqn:stress:exp}).
For general systems,
the stress tensor quantifies momentum flux densities
by virtue of the integral form of the continuity equation (\ref{eqn:flux}).
Momentum can flow through a system
either by motion of particles or
by the action of forces between adjacent portions of a system.

Ji and Liu~\cite{Ji:2021mfb} have argued against interpreting the stress tensor
of hadrons in terms of mechanical quantities like stress, pressure and shear,
and considered the ground state of the hydrogen atom~\cite{Ji:2022exr}
in their arguments against the stress interpretation.
In exploring the same system as Ji and Liu,
I have reached a different conclusion:
that the gravitational form factors and the stress tensor
quantify actual stresses acting inside the atom.

The electric field contribution to the stress tensor
can without question be interpreted as actual stresses,
since a static electric field does not carry momentum.
The electron and proton contributions are less straightforward,
and required more careful examination.
I concluded that they should be interpreted as actual stresses
on the basis that the proton and electron are stationary
(relative to the atomic barycenter)
in the hydrogen atom's ground state.
This is a generic feature of S-wave states in the pilot wave formulation
of quantum mechanics,
but the peculiar, strictly-tangential result for the
particle stress tensor (\ref{eqn:blt:ep})---which
is independent of the pilot wave interpretation---is suggestive
of this same feature.
If the proton and electron moved radially,
the radial momentum flux density should be non-zero;
and if the tangential momentum flux were due to circular motion
rather than stresses,
the orbital angular momentum would be non-zero.
The pilot wave interpretation readily accommodates the peculiar result
(\ref{eqn:blt:ep}),
which would seem paradoxical if we believed the stress tensor
described particle motion in this case.

Although I have reached different conclusions than Ji and Liu
regarding the interpretation of the
stress tensor, both their results and mine are
at odds with a common claim about mechanical properties of hadrons.
It is often speculated that mechanically stable systems must obey $D(0) < 0$.
Ji and Liu found---and I verified---that $D(0) > 0$ for the hydrogen atom
in its ground state.
Since the hydrogen atom's ground state is a paradigmatic example of a stable system,
the conjectured stability criterion could not be more soundly refuted\footnote{
  It should be remarked that the electron~\cite{Metz:2021lqv}
  and photon~\cite{Freese:2022ibw} in quantum electrodynamics also have $D(0) > 0$,
  and are also counterexamples to the stability condition.
}.

A major part of the rationale for assuming $D(0) < 0$ for stable systems
is given by Polyakov and Schweitzer~\cite{Polyakov:2018zvc}.
First of all, the isotropic pressure---after summed over
all constituents---obeys the von Laue condition~\cite{Laue:1911emt}:
\begin{align}
  \int \d^3 \bm{b} \,
  \mathfrak{p}^{(\mathrm{iso})}(\bm{b})
  =
  0
  \,.
\end{align}
If the isotropic pressure is compressive
(positive---corresponding to pushing)
near the system's center
and tensile
(negative---corresponding to pulling)
far from the center,
then mean-squared isotropic pressure radius
\begin{align}
  \int \d^3 \bm{b} \,
  \bm{b}^2
  \mathfrak{p}^{(\mathrm{iso})}(\bm{b})
  =
  \frac{1}{M}
  D(0)
\end{align}
would be negative.
Many authors assume that this ordering
of short-distance compressive stress and
long-distance tensile stress must hold for any stable system,
motivated by analogies
liquid drops~\cite{Polyakov:2018zvc}
or compact stars~\cite{Lorce:2018egm}.
However, one of the simplest and most iconic quantum systems---the
hydrogen atom---provides a stark counterexample
to exactly this ordering.
As shown in the left panel of Fig.~\ref{fig:isoani},
the stress in the hydrogen atom---which is dominated by the static
electric field---is negative at short distances and positive
at long distances.
This result does not undermine a mechanical
understanding of the hydrogen atom's stress tensor,
but merely provides a counterexample to a common assumption
about mechanical stability.

This work has largely been interpretive,
but an interesting finding of possibly practical significance
was given in Eq.~(\ref{eqn:cbar:force}).
For two-body non-relativistic systems,
if the form factor $\bar{c}_a(\bd^2)$ and the probability distribution
of particle $a$ were empirically measured,
Eq.~(\ref{eqn:cbar:force}) could be used to reconstruct the force law
binding the system.
It is currently unclear whether a generalization to many-body
or to relativistic systems exists,
and the quest for such generalizations will be the focus of follow-up research.
If this relation does generalize to many-body relativistic systems,
this offers an exciting opportunity to probe the force law
binding the proton or the pion
through the form factor $\bar{c}_q(\bd^2)$,
and to obtain a sharper understanding of quantum chromodynamics.

%%%%%%%%%%%%%%%%%%%%%%%%%%%%%%%%%%%%%%%%

\begin{acknowledgments}
  I gratefully acknowledge helpful discussions with
  Ian Clo\"{e}t, Joe Karpie, Gabriel Santiago and Christian Weiss.
  The research in this work received inspiration from the goals of the
  Quark Gluon Tomography Topical Collaboration of the U.S.\ Department of Energy.
  This work is supported by the Scientific Discovery through Advanced Computing
  (SciDAC) program in the Office of Science at the U.S. Department of Energy under
  contract DE-AC05-06OR23177 in collaboration with Jefferson Lab.
\end{acknowledgments}

%%%%%%%%%%%%%%%%%%%%%%%%%%%%%%%%%%%%%%%%%%%%%%%%%%%%%%%%%%%%%%%%%%%%%%%%%%%%%%%%

%\appendix

%%%%%%%%%%%%%%%%%%%%%%%%%%%%%%%%%%%%%%%%%%%%%%%%%%%%%%%%%%%%%%%%%%%%%%%%%%%%%%%%

\bibliography{references.bib}

\end{document}